\documentclass{emulateapj}
\usepackage{graphicx}
\usepackage[space]{grffile}
\usepackage{latexsym}
\usepackage{textcomp}
\usepackage{longtable}
\usepackage{multirow,booktabs}
\usepackage{amsfonts,amsmath,amssymb}
\usepackage{url}
\newif\iflatexml\latexmlfalse
\usepackage[utf8]{inputenc}
\usepackage[ngerman,english]{babel}

\def\rsun{\ifmmode {\rm R}_{\mathord\odot}\else $R_{\mathord\odot}$\fi}
\def\msun{\ifmmode {\rm M}_{\mathord\odot}\else $M_{\mathord\odot}$\fi}
\def\lsun{\ifmmode {\rm L}_{\mathord\odot}\else $L_{\mathord\odot}$\fi}
\def\kms{\ifmmode {\rm km~s}^{-1} \else km~s$^{-1}$\fi}

\shorttitle{An Exploration of the Statistical Signatures of Stellar Feedback}
\shortauthors{Boyden et al.}

\begin{document}


\title{An Exploration of the Statistical Signatures of Stellar Feedback}


\author{Ryan D. Boyden$^1$}
\author{Eric W. Koch$^2$}
\author{Erik W. Rosolowsky$^2$}
\author{Stella S. R. Offner$^{1,\dagger}$}
\affil{$^1$Department of Astronomy, University of Massachusetts, Amherst, MA 01003 USA;\\
$^2$Department of Physics, University of Alberta, Edmonton, T6G 2E1, Canada}
\email{$^\dagger$ soffner@astro.umass.edu}





\bibliographystyle{apj}

\begin{abstract}
All molecular clouds are observed to be turbulent, but the origin, means of sustenance, and evolution of the turbulence remain debated. One possibility is that stellar feedback injects enough energy into the cloud to drive observed motions on parsec scales. Recent numerical studies of molecular clouds have found that feedback from stars, such as protostellar outflows and winds, injects energy and impacts turbulence. We expand upon these studies by analyzing magnetohydrodynamic simulations of molecular clouds, including stellar winds, with a range of  stellar mass-loss rates and magnetic field strengths. We generate synthetic $^{12}$CO(1-0) maps assuming that the simulations are at the distance of the nearby Perseus molecular cloud. By comparing the outputs from different initial conditions and evolutionary times, we identify differences in the synthetic observations and characterize these using common astrostatistics. We quantify the different statistical responses using a variety of metrics proposed in the literature. We find that multiple astrostatistics, including principal component analysis, the spectral correlation function and the velocity coordinate spectrum, are sensitive to changes in stellar mass-loss rates and/or time evolution. A few statistics, including the Cramer statistic and velocity coordinate spectrum, are sensitive to the magnetic field strength. These findings demonstrate that stellar feedback influences molecular cloud turbulence and can be identified and quantified observationally using such statistics.
\end{abstract}

\section{Introduction}

Turbulence in the interstellar medium is ubiquitous and self-similar across many orders of magnitude \citep[]{brandenburg13}. Within molecular clouds, turbulence appears to play an essential role in the star formation process, regulating the efficiency at which stars form, seeding filaments and over-densities, and even potentially setting the stellar initial mass function  \citep{padoan12,federrath12,padoan14ppvi,hennebelle08,hopkins12,Offner14ppvi}. While the presence of supersonic motions is readily verified and has been studied using molecular spectral lines for several decades \citep{larson81},  the origin, energy injection scale, means of sustenance, and rate of dissipation remain debated. Moreover, molecular clouds display significant variation in bulk properties, ongoing star formation, and morphology. Consequently, it seems highly likely that these differences impact the turbulent properties of the gas and leave signatures --- \selectlanguage{english}but if so they are difficult to identify observationally.

Detailed study of the turbulence within molecular clouds is confounded by a variety of factors including observational resolution, projection effects, complex gas chemistry, and variable local conditions \citep[e.g.,][and references therein]{beaumont13}.  Any one molecular tracer only samples a limited set of gas densities and scales, so that reconstructing cloud kinematics reliably involves assembling a variety of tracers across different densities and scales \citep[e.g.,][]{gaches15}.
Many studies of cloud structure instead rely on a single gas tracer like CO, which is bright and exhibits widespread emission that reflects the underlying H$_2$ distribution \citep{bolatto13,heyer15}. Connecting such emission data to underlying turbulence and bulk cloud properties, however, is non-trivial. A variety of statistics have been proposed throughout the literature to characterize spectral data cubes and distill the complex emission information into more manageable 1D or 2D forms \citep[e.g.,][]{heyer97,rosolowsky99,rosolowsky08,burkhart09}. However, in most cases the utility of the statistic and its interpretation are not well constrained.

Numerical simulations, which supply full 6D information $(x,y,z, v_x, v_y, v_z)$, provide a means to study turbulence and constrain cloud properties. Prior studies have investigated how the turbulent power spectrum, inertial driving range, and fraction of compressive motions have influenced star formation \citep{klessen01,bate09b,federrath10}.  Other studies have connected simulated turbulent properties to observables such as  CO emission by performing radiative post-processing  \citep[e.g.,][]{padoan01,beaumont13,bertram14}. In some cases, this procedure is able to identify theoretical models that have good agreement with a given observation. Consequently, the most effective way to study turbulence in molecular clouds is by comparing observations with ``synthetic observations", in which the emission from the simulated gas is calculated via radiative transfer post-processing \citep[e.g.,][]{Offner08b,goodman11,Offner12, bertram15a}.   Recently, \citet{yeremi14} and Koch et al.~(2016) performed parameter studies of magneto-hydrodynamic (MHD) simulations in order to assess the sensitivity of common astrostatistics to changes in cloud velocity dispersion, virial parameter, driving scale, and magnetic field strength. They found that some statistics were responsive to changes in the temperature, virial parameter, Mach number and inertial driving range. These encouraging results raise the possibility that certain statistics may also be sensitive to energy input and environmental variation due to ongoing star formation and star formation feedback.

\subsection{Overview of Prior Statistical Studies and Feedback}

One fundamental puzzle in star formation is why the efficiency at which dense gas forms stars is only a few percent per free fall time \citep[][]{krumholz14review}.  Early three-dimensional hydrodynamic simulations demonstrated that supersonic turbulence decays rapidly and predicted that without additional energy input turbulence should decay significantly within a dynamical time \citep{stone98,maclow99}.  This implies that gravity should be able to efficiently form stars after a dynamical time. However, turbulence observed within molecular clouds does not appear to weaken and star formation efficiencies are small after several dynamical times \citep{KandT07}. One explanation for the longevity of observed turbulence is that motions are driven internally via feedback from forming or evolved stars \citep[][and references therein]{krumholz14ppvi}. In principle this should introduce a characteristic energy input scale \citep{carroll09,hansen12,Offner_2015}, which should impact turbulent statistics. However, from an observational prospective, stellar feedback is {\it messy} and identifying clear feedback signatures is complex for the reasons mentioned above. 
Disentangling feedback signatures from the turbulent background and assessing their impact is challenging since any low-velocity motions excited by feedback are often lost in the general cloud turbulence \citep{swift08,arce10,arce11}. 

Few prior numerical or observational studies have examined the response of turbulent statistics to stellar feedback. Several studies of the most commonly computed turbulent statistic, the velocity power spectrum, find that it may be sensitive to feedback. In numerical simulations, turbulence shaped by both isolated and clustered outflows exhibits a steepened velocity power spectrum \citep{nakamura07,cunningham09,carroll09}.  In observations of NGC 1333, \citet{swift08} identified a break in the power spectrum of the $^{13}$CO intensity moment map, which they attributed to a characteristic scale associated with the embedded protostellar outflows (the break is absent in the $^{12}$CO data). \citet{brunt09} and \citet{padoan09} reexamined the NGC1333 spectral cubes using principal component analysis (PCA) and the velocity coordinate spectrum (VCS) method, respectively, but found no evidence of outflow driving and concluded that the turbulence is instead predominantly driven on large scales. Numerical simulations of point-source (supernovae) driving also discovered changes in the spectral slope but found no obvious critical injection scale \citep{joung06}. 

Probability distribution functions (PDFs) of densities, intensities or velocities are also commonly computed \citep[e.g.,][]{nordlund99,lombardi06,federrath08}. Both observations and simulations suggest that gravity shapes the distribution at high densities   \citep{kainulainen09,collins12,girichidis14}, but the impact of feedback on PDFs is less clear. \citet{beaumont13} showed that observed CO velocity distributions extend to higher velocities than synthetic observations of simulations containing pure large-scale turbulence and gravity; they attribute this difference to expanding shells associated with stellar winds. \citet{Offner_2015} confirmed that when winds are included in simulations a high-velocity tail appears. In contrast, the column density probability distribution does not appear sensitive to the inclusion of stellar feedback \citep{beaumont13}.

The impact of feedback on higher order statistics, such as PCA, the spectral correlation function (SCF), dendrograms, the bispectrum and many others, is even less well explored \citep{rosolowsky99,heyer97,rosolowsky08,burkhart09}.  \citet{burkhart10}, in analyzing HI maps of the Small Magellenic Cloud, noted the possible signature of supernovae on the bispectrum, which appears as break around $\sim$160 pc. If true, it seems likely that other forms of stellar feedback influence statistics and impact other higher order statistics as well. 

In this paper, we aim to extend the study by Koch et al.~(2016, henceforth K16) by applying a suite of turbulent statistics to simulations with feedback from stellar winds. The simulated stellar winds produce parsec scale features and excite motions of several $\kms$ as a result of their expansion \citep{Offner_2015}. While protostellar outflows may also leave imprints in the turbulent distribution, winds appear to inject more energy on larger scales, which leaves a more distinct imprint on the gas velocity distribution \citep{arce11}.  By performing the analysis on synthetic CO spectral cubes, we aim to identify discriminating statistical diagnostics to apply to observed clouds that can pinpoint and constrain feedback: a ``smoking gun". 

In, \S\ref{methods} we describe the numerical simulations, production of synthetic CO data cubes, and astrostatistical toolkit we apply. We examine the response of each statistic to the presence of stellar winds in \S\ref{comparisons}. In \S\ref{distance}, we compare changes in the statistics between all pairs of outputs and assess the sensitivity to mass-loss rate, evolutionary time, and magnetic field strength. 
We discuss the results in \S\ref{discussion} and summarize our conclusions in \S\ref{conclude}.

\section{Methods}\label{methods}

\subsection{Numerical Simulations}

In this paper, we analyze the MHD simulations performed by \citet[henceforth OA15]{Offner_2015} of a small group of wind-launching stars  embedded in a turbulent molecular cloud. We refer the reader to that paper for full numerical details.  In brief, the calculations are performed using the {\sc orion} adaptive mesh refinement code \citep[e.g.,][]{li12}. They include supersonic turbulence, magnetic fields, and five star particles endowed with a prescription for launching isotropic stellar winds.  The domain size for all runs is 5 pc and the molecular gas is initially 10 K with a 3D velocity dispersion of $2.0~\kms$. The turbulent realization, magnetic field strength and wind properties vary between runs as stated in Table \ref{simprop}. We choose snapshots at different times from the various runs, including outputs at $t=0$, which contain turbulence unaffected by winds. Table \ref{simprop} summarizes the simulation properties for the specific evolutionary times we analyze here.  The simulation initial conditions correspond to a 3D Mach number of $\mathcal{M} = \sqrt{3} \sigma_v/c_s = 10.6$, virial parameter of $\alpha=5 \sigma_v^2 L/ (2 G M) = 1.0$, and plasma beta parameter ranging from $\beta=8 \pi M c_s^2/(L^3 B^2) = 0.02-0.6$, where $\sigma_v$ is the 1D velocity dispersion, $M$ is the cloud mass, $L$ is the cloud size, $c_s$ is the sound speed, and $B$ is the magnetic field strength. 


\begin{deluxetable}{lccc}
\tablecaption{Model Properties\tablenotemark{a} \label{simprop}}
\tablehead{ \colhead{Model } &  
   \colhead{B($\mu G$)} &
   \colhead{$\dot M_{\rm tot}(10^{-6 }\msun {\rm yr}^{-1})$\tablenotemark{b}} &
   \colhead{$t_{\rm run}$(Myr)}}
\startdata
W1T1t0.1  & 13.5 &  41.7 & 0.1 \\
W1T2t0.1  & 13.5 &  41.7 & 0.1 \\
W1T2t0.2  & 13.5 &  41.7 & 0.2 \\
T2t0  & 13.5 &  ... & 0 \\
W2T2t0.1  & 13.5 &  4.5 & 0.1 \\
W2T2t0.2  & 13.5 &  4.5 & 0.2 \\
T3t0  & 5.6 &  ... & 0 \\
W2T3t0.1  & 5.6 &  4.5 & 0.1 \\
T4t0  & 30.1 &  ... & 0 \\
W2T4t0.1  & 30.1 &  4.5 & 0.1 
\enddata
\tablenotetext{a}{Model name, initial mean magnetic field, the total stellar mass-loss rate, the evolutionary time. All models have $L=5$pc, $M=3762 \msun$, $T_i=10$K and $N_*=5$.   } 
\tablenotetext{b}{The estimated mass-loss rate from all stellar winds in Perseus is $9.49 \times 10^{-6}\msun$yr$^{-1}$ (A11). }
\end{deluxetable}

\subsection{CO Emission Modeling}

Following OA15,  we post-process each output with the radiative transfer code {\sc radmc-3d}\footnote{\url{http://www.ita.uni-heidelberg.de/~dullemond/software/radmc-3d/}} in order to compute the $^{12}$CO (1-0) emission. We solve the equations of radiative statistical equilibrium using the Large Velocity Gradient (LVG) approach \citep{shetty11}. We perform the radiative transfer using the densities, temperatures, and velocities of the simulations on a uniform $256^3$ grid.  This is the simulation basegrid resolution, which is conservatively updated using information from the finer AMR levels \citep{li12}. OA15 compare several different quantities and statistics for a simulation evolved with and without additional AMR levels and for different flattened grid sizes. They find little difference, so we expect our conclusions from the CO modeling to be similar for larger grids. 

We convert to CO number density by defining $n_{\rm H_2} = \rho/(2.8 m_p)$ and adopting a CO abundance of [$^{12}$CO/H$_2$] =$10^{-4}$ \citep{frerking82}.  Gas above 800 K or with $n_{\rm H_2} < 10$ cm$^{-3}$ is set to a CO  abundance of zero. This effectively means that gas inside the wind bubbles is CO-dark. The CO abundance in regions with densities $n_{\rm H_2} > 2 \times 10^4$ cm$^{-3}$ is also set to zero, since CO freezes-out onto dust grains at higher densities \citep{Tafalla_2004}.  Some CO may remain above this threshold \citep{hocuk15}. However, in the strongest wind case (W1T2t0.2), which has the most gas compression, only 0.035\% of the volume contains densities in excess of this value, so we expect the choice of freeze-out cutoff to have minimal impact on the statistics. In the radiative transfer calculation, we include sub-grid turbulent line broadening by setting a constant micro-turbulence of 0.25 $\kms$. The data cubes have a velocity range of $\pm 20~\kms$ and a spectral resolution of $\Delta v = 0.156~ \kms$. 

To mimic the effects of observational noise, we add Gaussian noise with a standard deviation of $\sigma_{\rm rms} = 0.1$K. This is comparable to the noise in the FCRAO $^{12}$CO COMPLETE survey of local star forming regions \citep{ridge06}.
  
We produce synthetic observations of the nearby Perseus molecular cloud by setting the spectral cubes at a distance of 250 pc. The emission units are converted to temperature (K) using the Rayeigh-Jeans approximation. 

 In order to assess the impact of the radiative transfer on the statistics, we also construct position-position-velocity (PPV) cubes using the raw simulation density and velocity. Instead of CO emission, each PPV cube voxel contains the total mass along the line-of-sight with velocities contained in a given velocity channel range. These cubes are constructed using the same spatial (256$^3$) and velocity resolution ($\Delta v = 0.156~ \kms$) as the CO spectral cubes. 
These simpler cubes eliminate the effects of excitation variations and optical depth. We present the analysis of these cubes in the Appendix and discuss the implications in \S\ref{discussion}.


\subsection{Statistical Toolkit}\label{toolkit}

We perform the statistical analysis using {\sc TurbuStat}\footnote{\url{http://turbustat.readthedocs.org/en/latest/} }, a Python package developed by K16 that contains 16 turbulent statistics culled from the literature.  Table \ref{stats} summarizes the statistics contained in this astrostatistical toolkit that we consider here.  K16 provide a detailed description of each turbulent statistic, and so we give only a brief overview here.   After calculating the statistic for each cube {\sc TurbuStat} measures differences between spectral cubes by computing a pseudo-distance metric as first proposed in \citet{yeremi14}.  We briefly describe the distance definitions in each section of \S\ref{comparisons} and refer the reader to K16 for the corresponding mathematical formulae.

Below we group the statistics into three categories based on their method of analysis: {\it intensity statistics} quantify emission distributions, {\it Fourier statistics} analyze N-dimensional power spectra obtained through spatial integration techniques, and {\it morphology statistics} characterize the structure of the emission.  One essential property of a ``good" statistic is that it is insensitive to turbulent seed or viewing angle. K16 run the fiducial case for five different random seeds. They evaluate each statistic and find that all but modified centroid analysis (MVC) is insensitive to the initial seed. Thus,  we focus on the statistical formulations that exhibit meaningful responses to changes in underlying physical parameters, and we exclude MVC from our analysis. 



This paper extends the analysis of the K16 study by examining simulations including feedback from stellar winds and considering larger grid resolutions.  However, our simulation suite does not utilize experimental design to set the parameter values.  \citet{Yeremi_2014} cautioned that comparisons between outputs in one-factor-at-a-time approaches may give a misleading signal since the statistical effects are not fully calibrated. 

\begin{deluxetable*}{llll}
\tablecaption{ Statistics\tablenotemark{a} \label{stats}}
\tablehead{ \colhead{Family } & \colhead{Name} & \colhead{Comparison Metric\tablenotemark{c}}  & \colhead{Citations\tablenotemark{b}}}
\startdata
 		& Probability Distribution Function (PDF) & Histogram$^{\rm 2D}$ & \citet{nordlund99} \\
	& PDF Skewness					& Histogram$^{\rm 2D}$  & \citet{kowal07,burkhart09} \\
Intensity 	& PDF Kurtosis 					& Histogram$^{\rm 2D}$  & \citet{kowal07,burkhart09} \\
Statistics  		& Principal Component Analysis (PCA)	& Eigenvalues & \citet{heyer97,brunt02a, brunt02b} \\
		 & Spectral Correlation Function (SCF) 	& Surface & \citet{rosolowsky99,padoan99} \\
		 & Cramer & Distance & \citet{yeremi14} \\ \hline
  		&  Spatial Power Spectrum (SPS) & Power-law Slope$^{\rm 2D}$  & \citet{lazarianp04} \\
     		&  Velocity Channel Analysis (VCA) & Power-law Slope & \citet{lazarianp00,lazarianp04}\\ 
Fourier    & Velocity Coordinate Spectrum (VCS) & Power-law Slope & \citet{kowal07,chepurnov09}\\ 
Statistics 		& Bispectrum & Bicoherence Matrix$^{\rm 2D}$ & \citet{burkhart09,burkhart10} \\ 
		& $\Delta$-Variance & Spline Fit$^{\rm 2D}$ & \citet{stutzki98,ossenkopf08a,ossenkopf08b} \\
		& Wavelet Transform & Power-law Slope$^{\rm 2D}$ &  \citet{gill90}\\  \hline
		&  Genus & Spline Fit$^{\rm 2D}$  & \citet{gott86,kowal07}\\  
Morphology& Dendrogram Leaves & Power-law Slope & \citet{rosolowsky08,goodman09} \\  
Statistics	 & Dendrogram Feature Number & Histogram & \citet{burkhart13b}
\enddata
\tablenotetext{a}{List of all statistics we calculate. } 
\tablenotetext{b}{A list of seminal papers that have either developed or explored this statistic in detail in the context of molecular clouds.}
\tablenotetext{c}{The form of the pseudo-distance metric used to assess the degree of difference between two datacubes. The mathematical definition for each is given in K16.}
\tablenotetext{2D}{ Statistics that are performed using the 2D integrated emission rather than the full 3D spectral cube.}
\end{deluxetable*}

\section{Statistical Comparisons}\label{comparisons}

In \S\ref{distance} we calculate the distance between each pair of spectral cubes for each statistic. However, since few prior statistical studies have studied the impact of feedback, we first investigate and present the statistical response to feedback for two fiducial outputs: W1T2t0.2 and T2t0.
T2t0 is a simulated turbulent molecular cloud (turbulent realization T2) prior to wind launching. W1T2t0.2 begins with the same turbulence as T2t0 (T2), but it has evolved for 0.2 Myr (t0.2) with wind launching model W1. 
Thus, the two runs begin with the same turbulent seed, but the turbulence in one run is shaped by feedback, while in the other the turbulence is ``pristine". For each statistic in Table \ref{stats}, we compare the results produced with runs W1T2t0.2 and T2t0 and identify qualitative differences. 

 We restrict the comparison to views along the $z-$direction. However, we expect statistically similar distributions for other views since we confine our study to those statistics K16 demonstrated to be insensitive to the turbulent seed. Consequently, large distances reflect real changes between T2t0 and W1T2t0.2 and are not produced by random variations in the underlying turbulence, i.e., outputs T2t0 and T2t0.2 (identical physical parameters at a different time without winds) would be statistically indistinguishable.

\subsection{Intensity Statistics}\label{comparisons_inten}

In this section we discuss statistics that quantify the emission distribution: the probability distribution function (PDF), skewness, kurtosis, principal component analysis (PCA), and the spectral correlation function (SCF). 
We also compute the Cramer statistic, but as a one-point statistic, it directly defines a distance, so we defer its discussion until $\S4$.


\subsubsection{Probability Distribution Function}

\begin{figure}[h!]
\begin{center}
\includegraphics[width=1\columnwidth]{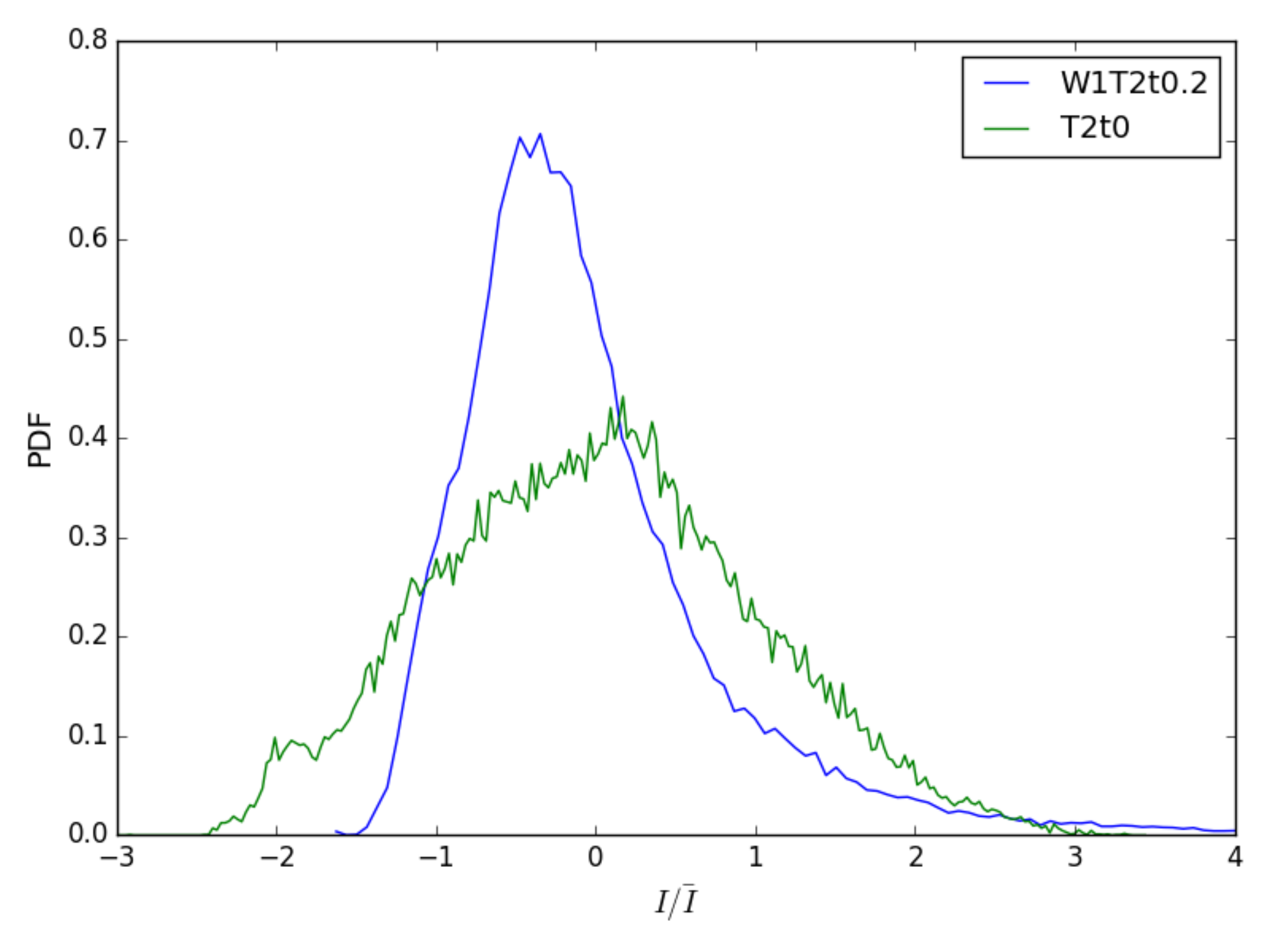}
\caption{PDFs of the integrated intensity moment maps for runs W1T2t0.2 (blue) and T2t0 (green). The integrated intensities are normalized to have a mean of 0 and standard deviation of 1.%
\label{pdf} }\selectlanguage{english}
\end{center}
\end{figure}

We calculate the probability distribution function (PDF) of the normalized integrated intensity maps. The intensity in each pixel is weighted by the respective error, $1/\sigma^2$, where $\sigma^2$ is proportional to the number of channels.
The PDF is simply a measure of the relative number and range of integrated intensities. Figure \ref{pdf} shows the two fiduical PDFs. Both runs exhibit Gaussian behavior, although the output with winds, W1T2t0.2, is more peaked around the mean and has a longer tail towards higher integrated intensities. 
These differences arise because the winds create shells with CO-brightened rims that produce higher intensities  than the compressions created by the strongest shocks in the case of pure turbulence. 

 The PDF distance metric is defined as the sum of the bin differences (according to the Hellinger distance formula) between the normalized PDFs. Under this definition, the large differences in the PDF breadth, which are visually apparent, create a large distance between the two distributions. 

Prior studies have shown that the widths of the density and column density PDFs increase with Mach number \citep[e.g.,][]{nordlund99,ostriker01}. In the strong wind case, the effective Mach number is about 10\% higher (OA15), however, this is not sufficient to explain the difference in Figure \ref{pdf}. The intensity distribution of the case with winds is broadened by a combination of higher densities and temperatures (the shells are warmer than the ambient turbulent gas),  which enhances the CO excitation.

\subsubsection{Skewness and Kurtosis}

\begin{figure}[h!]
\begin{center}
\includegraphics[width=1\columnwidth]{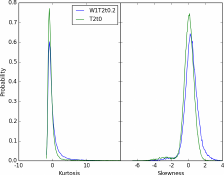}
\caption{Kurtosis (left) and skewness (right) PDFs for output W1T2t0.2 (Blue) and output T2t0 (green).%
\label{skew}}\selectlanguage{english}
\end{center}
\end{figure}

The skewness and kurtosis provide a means of classifying the shape of the intensity distribution. They are the third- and fourth-order statistical moments of the PDF, respectively. Skewness is a measure of the symmetry of the data distribution. PDFs that are symmetric around the center point have low skewness. If there is an excess of high values, the skewness will be positive, while an excess of low values produces negative skewness. Kurtosis quantifies the extent and ``peakiness" of the distribution. Normally distributed data has a kurtosis of zero, data more concentrated than a Gaussian will have negative kurtosis, and flatter data and data with an extended tail will exhibit positive kurtosis.   Following \citet{burkhart09}, we compute each higher-order moment within a small, circular region with a radius of five pixels in the integrated intensity map.  Figure \ref{skew} shows histograms of these moments as calculated from all circular patches.

The kurtosis PDFs exhibit similar behavior: both are centered at zero and sharply decrease with increasing kurtosis magnitude. However, the T2t0 distribution falls off more quickly than that of W1T2t0.2. This likely occurs because the winds generate a more extreme range of high intensity values; the intensity distribution deviates further from a normal distribution and exhibits a tail of high intensities.

The skewness PDFs have similar shapes, and both have a small tail at negative skewness. However, the W1T2t0.2 distribution center is shifted to positive skewness, while the T2t0 distribution is centered at zero. This makes sense since the winds in W1T2t0.2 create an excess of high-intensity values.

 The distance metrics for the skewness and kurtosis, like that of the PDF, are defined as the sum of the bin-wise differences computed using the Hellinger distance formula. Consequently, the disparate shapes of the normalized PDFs in Figure \ref{skew} produce significant distance between the outputs. 

Simulations of pure turbulence find that as the Mach number increases, the skewness and kurtosis of the column density PDF also increase \citep{kowal07,burkhart09}. Higher Mach number flows have stronger shocks, which increase the fraction of high-density, and hence high-column density, material. This is consistent with our results, since the winds create density enhancements and the CO intensity serves as a proxy for the gas column density.


\subsubsection{Principal Component Analysis}

Principal component analysis (PCA) determines a set of orthogonal axes that maximize the variance of the data. As applied to spectral data cubes, it identifies differences between the line profiles, and thus, is a useful tool for distinguishing between kinematic changes and noise \citep{heyer97}. Subsequent work established an empirical and analytic formalism connecting PCA to the underlying turbulent velocity fluctuations, including the spectral slope \citep{brunt02a,brunt02b,brunt13}. In PCA analysis, the first step involves constructing a 2D covariance matrix between the velocity channels of the data cube. Next, the eigenvalues and eigenvectors of this matrix are determined. Here, we use the magnitude of the eigenvalues to assess the degree of difference between two datasets. The relative magnitudes of the eigenvalues are a simple description of how the power in the data cube projects onto the linear PCA basis.

\begin{figure}[h!]
\begin{center}
\includegraphics[width=1\columnwidth]{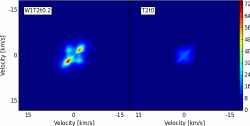}
\caption{Covariance matrices of the velocity channels for runs W1T2t0.2 (left) and T2t0 (right). The  axes indicate the two velocity channels in which we calculate the total covariance summed over all positions. The colorbar denotes the covariance magnitude.%
\label{pca1}}\selectlanguage{english}
\end{center}
\end{figure}

Figure \ref{pca1} shows the velocity channel covariance matrices of outputs W1T2t0.2 and T2t0. Both show a signal for velocities $|v| \lesssim 2~\kms$, which roughly encompasses the range of turbulent gas velocities. However, W1T2t0.2 exhibits multiple strong covariance peaks at velocities of a few $\kms$. These features exist to a lesser degree for T2t0, but feedback augments and further separates the peaks. The strongest covariance corresponds to the typical expansion rate of the wind shells, which is $\sim 1-3~\kms$.

\begin{figure}[h!]
\begin{center}
\includegraphics[width=1\columnwidth]{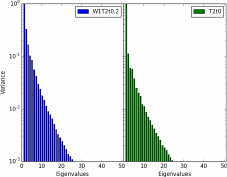}
\caption{The first 50 covariance matrix eigenvalues for runs W1T2t0.2 (left) and T2t0 (right). For each plot, the eigenvalues are normalized with respect to the maximum eigenvalue. Their magnitudes denote the relative variance described by that principal component.%
\label{eigenvalue}}\selectlanguage{english}
\end{center}
\end{figure}

Because the eigenvalues provide a measure of the strength of different eigenvectors, they also serve as an indirect measure of the amount of power on different scales \citep{brunt02a,brunt02b}. Figure \ref{eigenvalue} shows the relative sizes of the largest eigenvalues. Our algorithm calculates the first 50 and uses  the normalized sum of their differences to define the distance between two data cubes. However, as the figure shows, only the first ten are significant,  and these dominate the distance metric. For observations, the number of significant eigenvalues depends on the scale of the image, and usually those beyond the first 8-10 are dominated by noise and thus contain little information.  We find a clear difference in the dominant eigenvalues for the cases with and without feedback. The case with feedback has more significant second, third and fourth eigenvalues,  which increase the distance. This variation is likely due to the additional emission structure created by the winds. 

\subsubsection{Spectral Correlation Function}

The spectral correlation function (SCF) is the normalized root-mean-square difference between two spectra as a function of their projected separation \citep{rosolowsky99}. The SCF manifests as a power-law, where flatter slopes indicate more kinematic correlation across spatial scales (large hierarchical emission structures), while steeper slopes indicate less correlation between large and small scales (smaller discrete emission structures). The SCF serves as a useful comparison metric for both simulations and observations \citep{padoan99,yeremi14,gaches15}; however, no direct link between the SCF and turbulent properties has been formulated.

\begin{figure}[h!]
\begin{center}
\includegraphics[width=1\columnwidth]{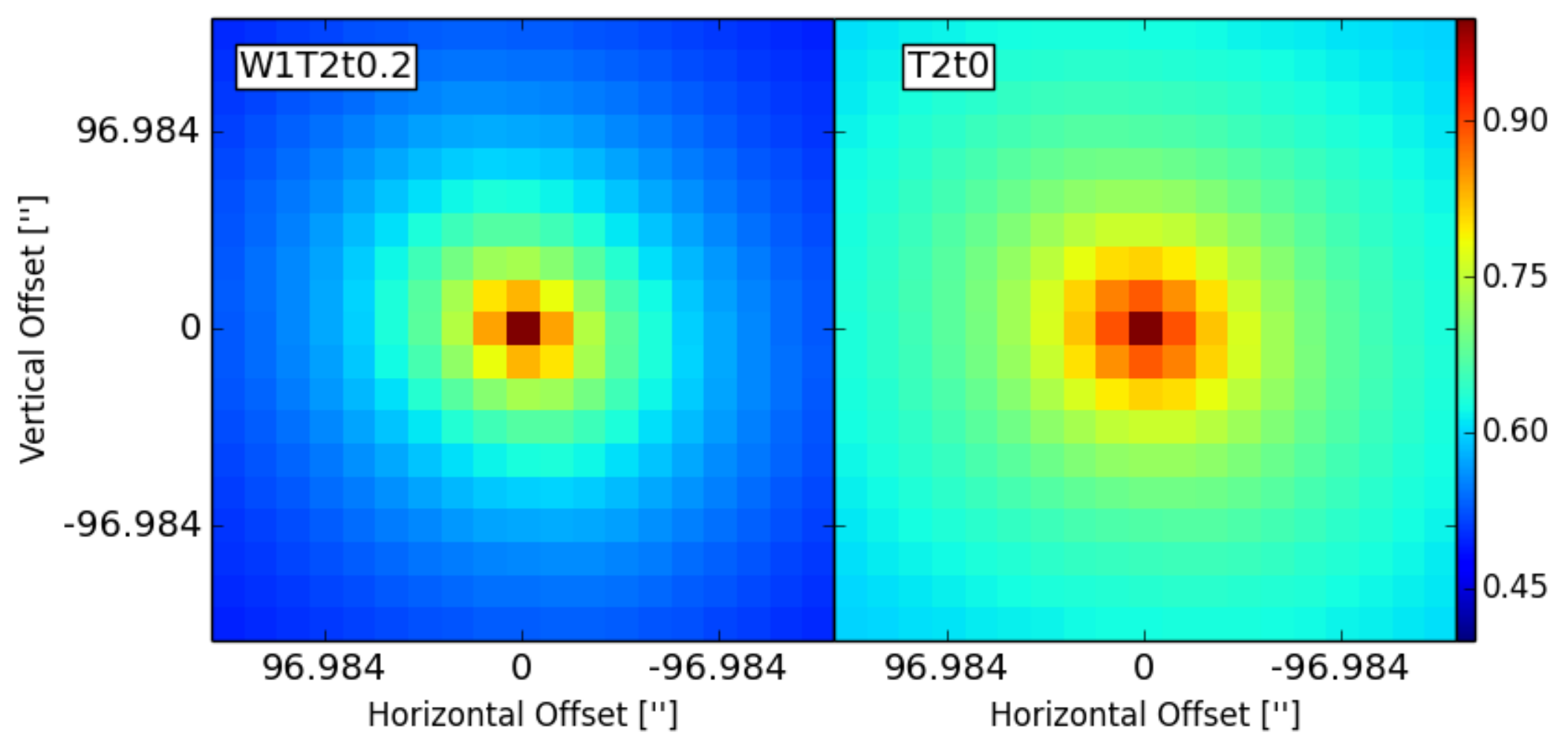}
\caption{SCF surface for outputs W1T2t0.2 and W2T2t0. The $x$ and $y$ axes denote the amount of horizontal and vertical offset, respectively, used in the SCF computation, and the colorbar denotes the SCF value. A value of 1 indicates complete correlation while a value of 0 denotes complete lack of  correlation.%
\label{scf} }\selectlanguage{english}
\end{center}
\end{figure}

\begin{figure}[h!]
\begin{center}
\includegraphics[width=1\columnwidth]{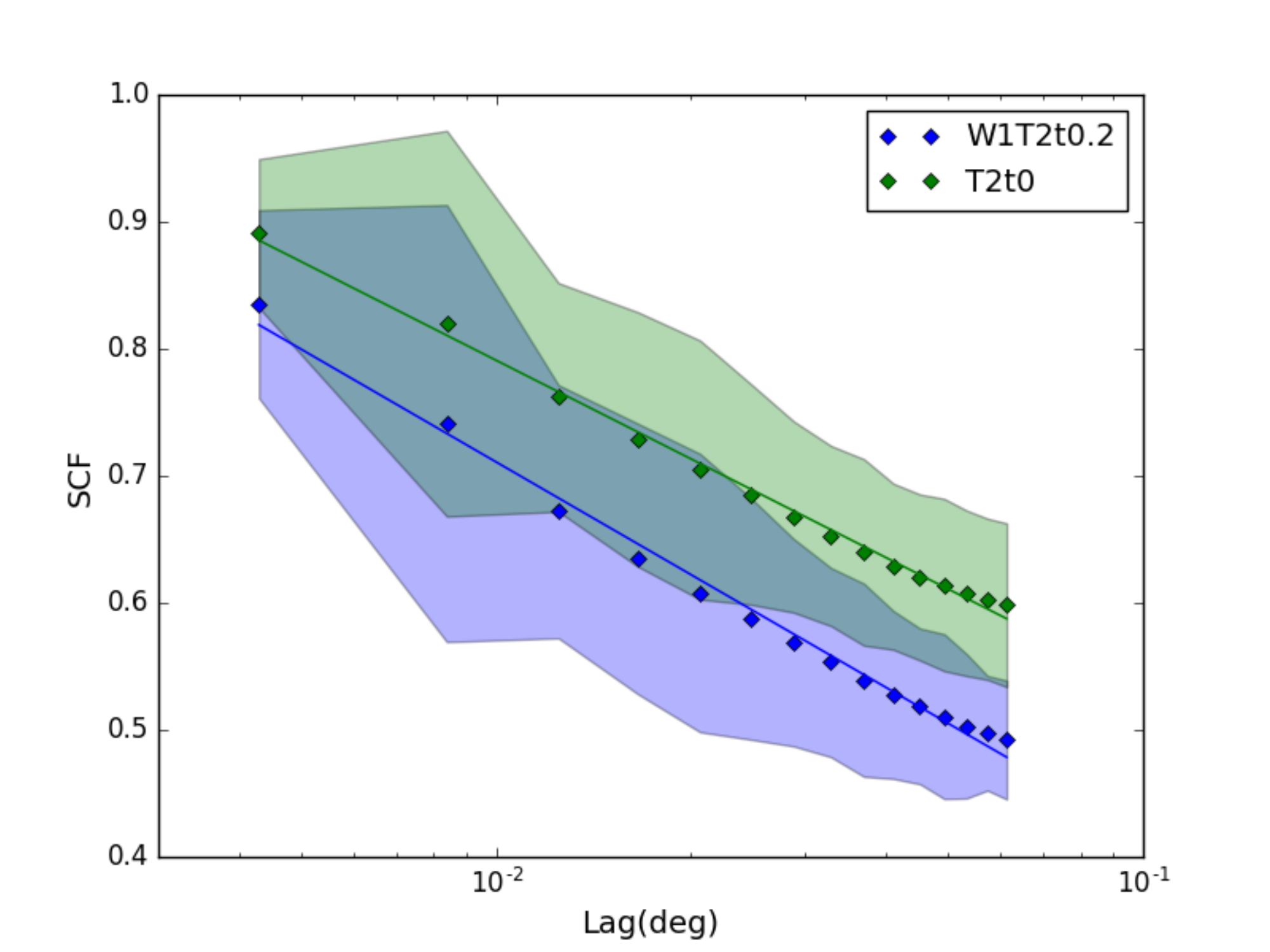}
\caption{SCF spectrum (azimuthal average of the SCF surface in Fig \ref{scf}) for outputs W1T2t0.2 and W2T2t0. The slopes are $-0.294 \pm 0.008$ and $-0.258\pm0.005$ for the wind and non-wind runs, respectively.  The shaded region indicates the standard deviation of the data in each bin.
\label{scfsp} }\selectlanguage{english}
\end{center}
\end{figure}

We calculate the SCFs of outputs W1T2t0.2 and T2t0 using an array of projected separations ranging from  0'' to 113". 
Figure \ref{scf} depicts the SCF surfaces of our two fiducial outputs. 
The SCF surface of W1T2t0.2 is peakier than that of T2t0, which corresponds to a steeper SCF spectrum slope as shown in Figure \ref{scfsp}.  The SCF spectrum is defined as an azimuthal average of the SCF surface over annuli of different radii or ``lag". 

 The SCF distance between two outputs is proportional to the sum of the differences between corresponding points in the SCF surface, weighted by the distance from the center. Thus, the variation in shape illustrated in Figure \ref{scfsp} enhances the distance. This shape variation is also encapsulated by the SCF spectrum, which presents an alternative means of comparison \citep{padoan99,gaches15}. 

 Both spectrum slopes are comparable to the SCF slope of -0.29 found by \citet{gaches15} for simulations of non-magnetized turbulence without feedback. However, \citet{gaches15} also employ chemical networks to model the abundance distribution of CO, which may account for the better agreement with the W1T2t0.2 slope.  Since the SCF spectrum exhibits no characteristic feature associated with feedback and the SCF slope also depends on resolution, we expect this statistic to be most effective when comparing different subregions within a cloud.  

\subsection{Fourier Statistics}  

In this section we present statistics based on a Fourier analysis of the spectral cube: velocity channel analysis (VCA), velocity coordinate spectrum (VCS), spatial power spectrum (SPS), bicoherence, $\Delta$-variance and wavelet transform. Since K16 was unable to identify a formulation of the MVC method that reliably discriminated between models, we do not include it here.

\subsubsection{Spatial Power Spectrum}\label{ps}

The Fourier power spectrum is one of the most widely computed turbulent statistics. Numerical simulations over the last decade have confirmed that the velocity power spectral slope in one dimension is $P_v(k) \propto k^{-2}$ for supersonically turbulent gas \citep[][and references therein]{maclow04,MandO07}. The slope is similar or slightly flatter for a magnetized gas where the gas and field are well-coupled. The power spectrum of the 3D density distribution of turbulent gas is $P_\rho(k) \propto k^{-1.5}$ and $k^{-2.3}$ for solenoidal and compressive driving, respectively \citep{federrath10}. Observationally, the situation is more complex since the intensity distribution in a spectral line cube is a product of both density and velocity fluctuations, which are inextricably entangled. For lower density tracers, like $^{12}$CO, the gas becomes optically thick and emission saturates along high-density sight-lines through the cloud. \citet{lazarianp04} predicted that the intensity power spectrum  intensity field follows $P(k) \propto k^{-11/3}$ and saturates at $P(k) \propto k^{-3}$ in the optically thick limit. This was confirmed in numerical simulations by \citet{burkhart13a}, who post-processed MHD simulations to produce synthetic CO maps in different optical depth regimes. Because the emission behaves differently in different optical depth limits, it is possible to probe the underlying density and velocity slopes by analyzing the spectrum of different slices within the spectral cube \citep{lazarianp00}, a technique that we discuss further in \S\ref{VCA}.

\begin{figure}[h!]
\begin{center}
\includegraphics[width=1\columnwidth]{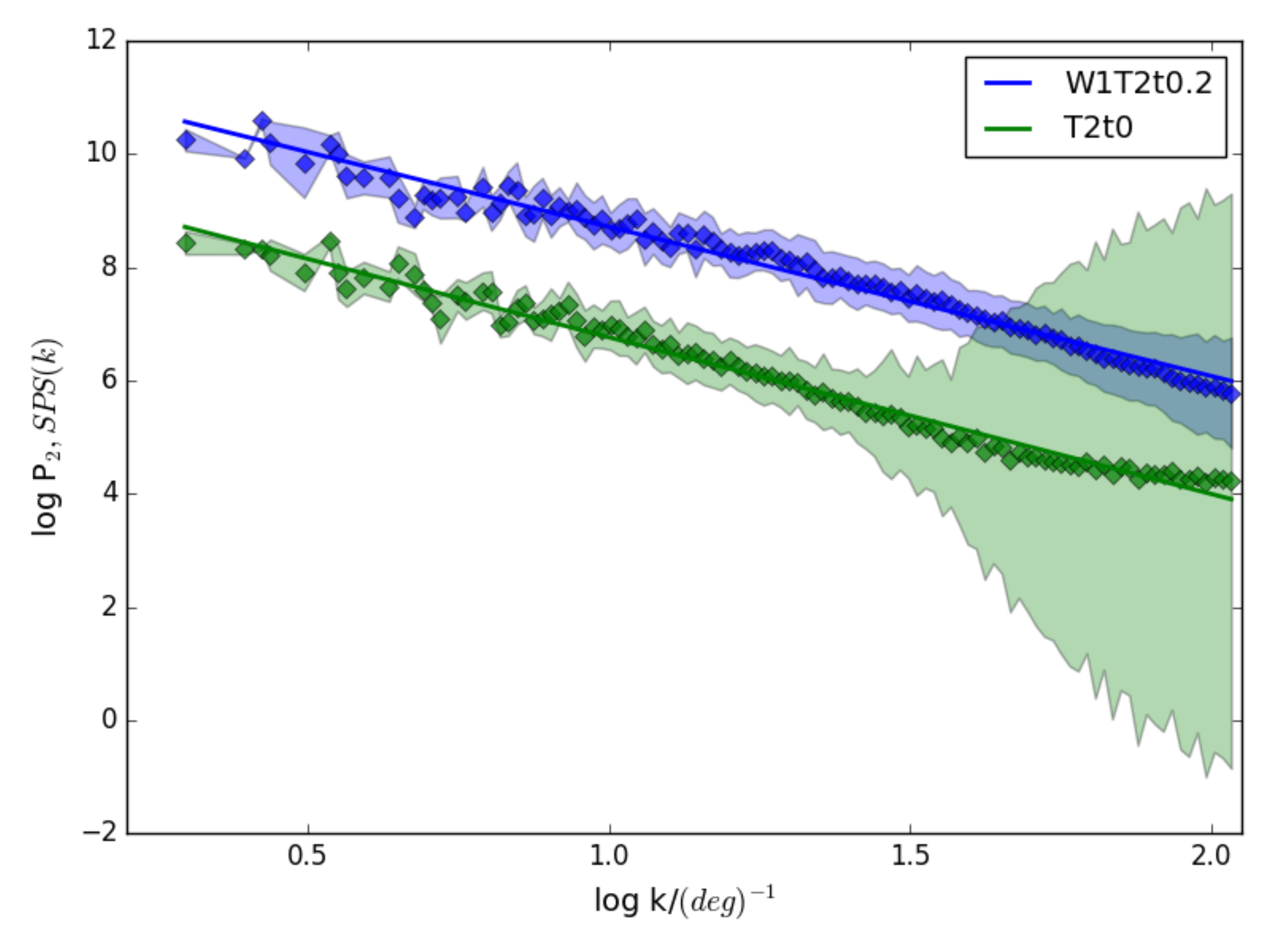}
\caption{\selectlanguage{english}SPS for outputs W1T2t0.2 (blue, top) and T2t0 (green, bottom), where the $x$-axis scale in wavenumber is normalized to units of deg$^{-1}$ to make it dimensionless.  The solid lines indicate the power-law fits. 
The line fits have slopes of -2.64$\pm$0.04 for W1T2t0.2 and -2.77$\pm$0.04 for T2t0. 
  The shaded region indicates the standard deviation of the data in each bin.
\label{psfig}}
\end{center}
\end{figure}

To obtain the SPS, we compute the Fourier transform of the integrated intensity map and calculate the 2D power spectra of the two-point autocorrelation function. We then radially average the power spectra over bins in spatial frequency. Fitted power laws for each 1D power spectrum are shown in Figure \ref{psfig}. We find 
a constant horizontal offset with the wind output exhibiting more power overall and a slightly flatter slope.   The SPS distance metric is defined according to a linear function of the difference between the two fitted slopes, weighed by the uncertainty of each slope added in quadrature. Formally, this is the t-statistic of the difference in the slopes. The similar slopes shown in Figure \ref{psfig} thus produce a relatively small distance.

The slope of the pure turbulence run, $-2.77 \pm 0.04$, is close to $k^{-3}$, which is predicted as the limiting slope for a turbulent optically thick gas \citep{lazarianp04,burkhart13a}.  We do not see any break representative of a characteristic driving scale. It is possible that the high gas optical depth may hide any underlying break in the spectrum. A different result could be expected for a more optically thin tracer such as $^{13}$CO \citep[e.g.,][]{swift08}, but OA15 do not find a quantitatively different result for $^{13}$CO for these simulations. Since the two outputs have the same average gas density, we might expect them to have the same SPS slope. However, the slightly flatter slope of W1T2t0.2 suggests that feedback and optical depth are somewhat degenerate in their impact of the SPS. 

\subsubsection{Velocity Channel Analysis and Velocity Coordinate Spectrum}\label{VCA}

Velocity Channel Analysis (VCA) and Velocity Coordinate Spectrum (VCS) are techniques that isolate how fluctuations in velocity contribute to differences between spectral cubes \cite[e.g.,][]{lazarianp00,lazarianp04}. 
VCA produces a 1D power spectrum as a function of spatial frequency, while VCS yields a 1D power spectrum as a function of velocity-channel frequency (velocity wavenumber). For outputs W1T2t0.2 and T2t0, we first compute the three-dimensional power spectrum. To obtain the VCA, we calculate a one-dimensional power spectrum by integrating the 3D power spectrum over the velocity channels and then radially averaging over the two-dimensional spatial frequencies. A portion of the resultant 1D power spectrum is then fit with a power law. For VCS, we reduce each 3D power spectrum to one dimension by averaging over the spatial frequencies. This yields two distinct power laws, which we fit individually using the segmented linear model described in K16. The fit at larger scales describes bulk gas velocity-dominated motion; the fit at smaller scales describes gas density-dominated motions \citep{chepurnov09}. \citet{kowal07} find that the density-dominated regime is sensitive to the magnetic field strength, where stronger fields correspond to steeper slopes.

\begin{figure}[h!]
\begin{center}
\includegraphics[width=1\columnwidth]{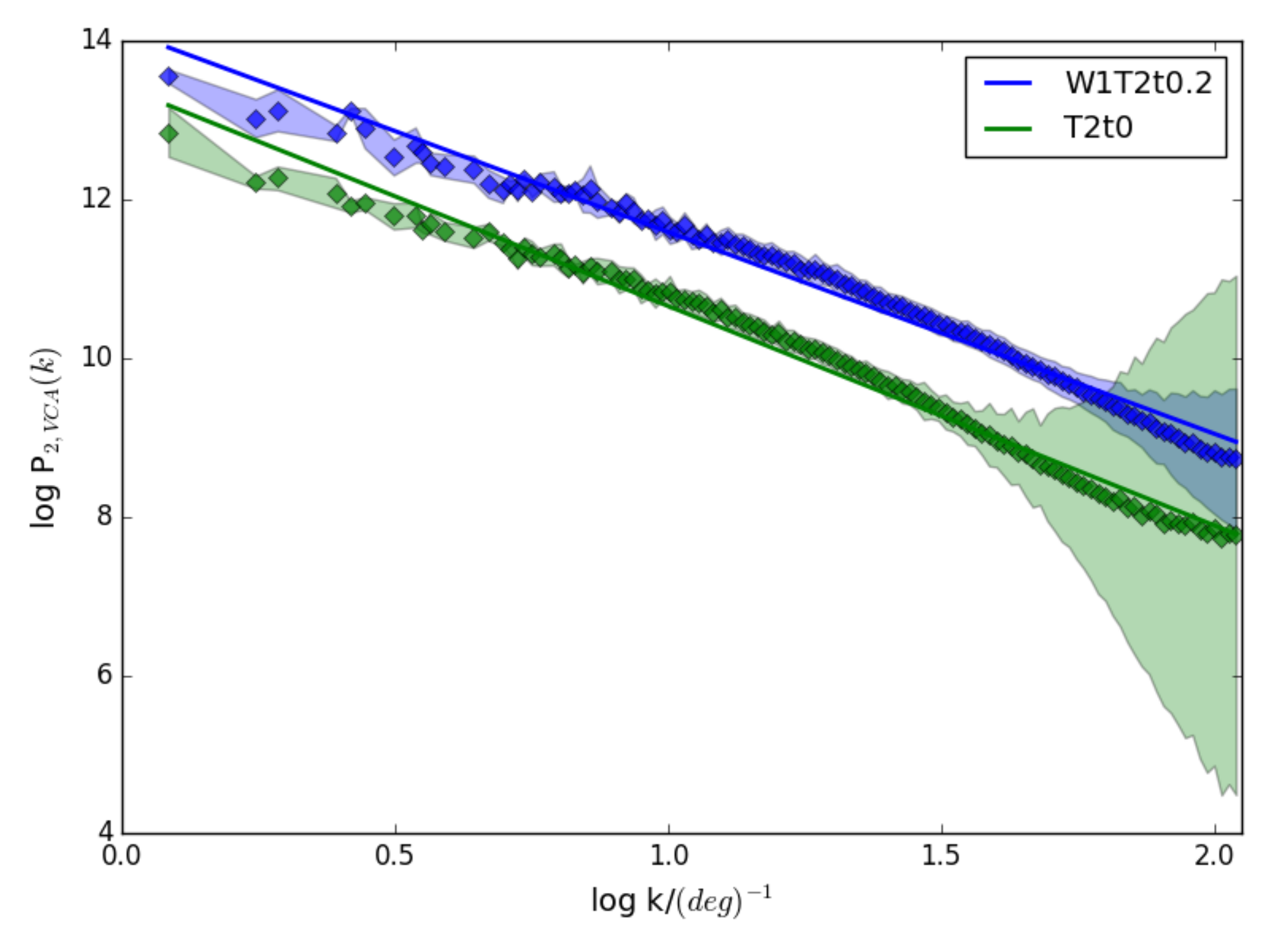}
\caption{\selectlanguage{english}VCA as a function of spatial frequency $k$ for outputs W1T2t0.2 (top, blue) and T2t0 (bottom, green), each fitted by a single power law.  The wavenumber is normalized to units of deg$^{-1}$ to make the $x-$axis dimensionless. We report a slope of -2.55$\pm$0.03 for W1T2t0 and -2.77$\pm$0.03 for T2t0. 
 The shaded region indicates the standard deviation of the data in each bin.
\label{vcafig}}
\end{center}
\end{figure}

\begin{figure}[h!]
\begin{center}
\includegraphics[width=1\columnwidth]{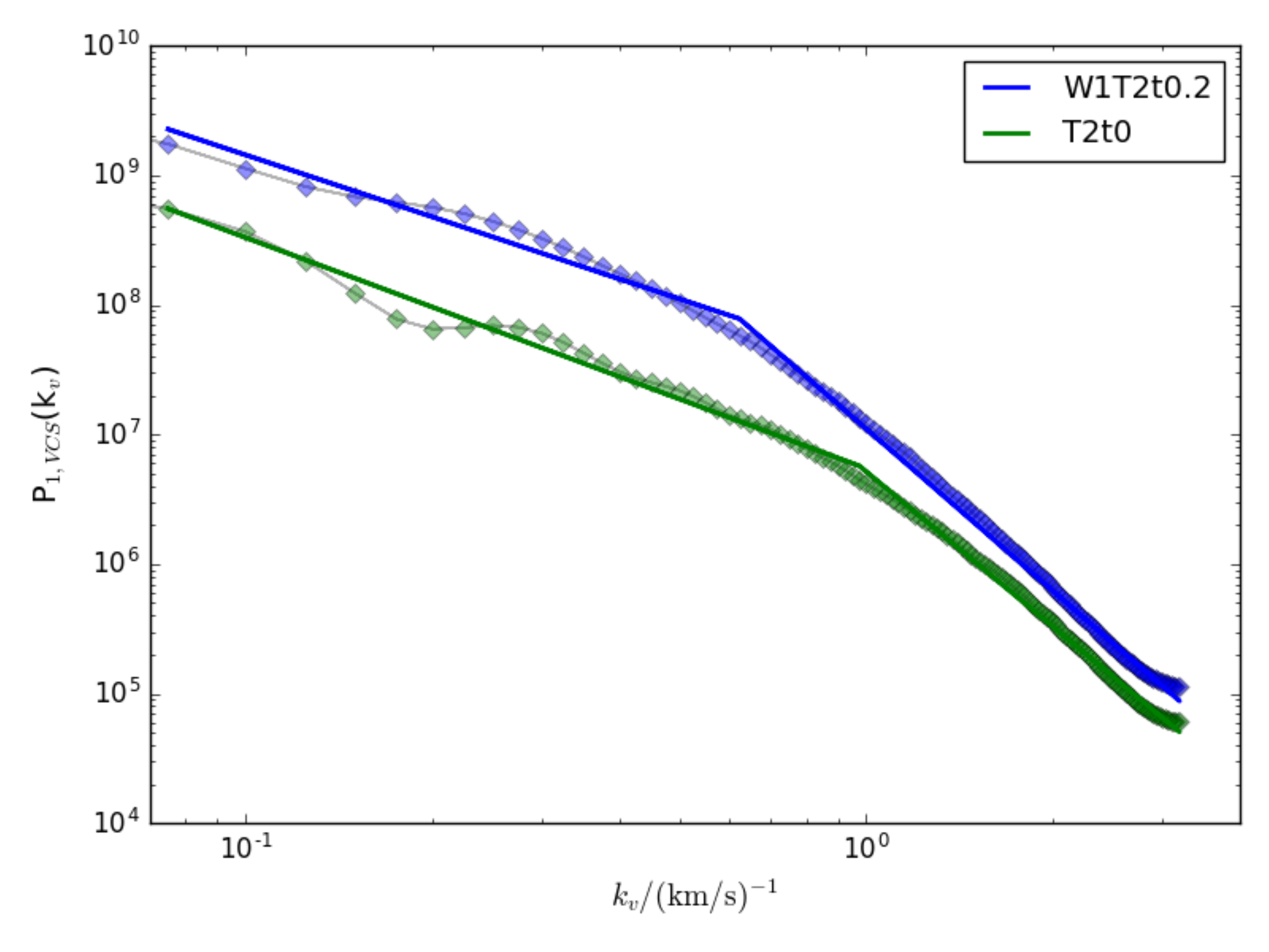}
\caption{\selectlanguage{english} VCS as a function of velocity-frequency for outputs W1T2t0.2 (top, blue) and T2t0 (bottom, red).  The wavenumber is normalized to units of (km/s)$^{-1}$ to make the $x-$axis dimensionless. The segmented power law fits are overlaid. For W2T2t0.2, we report slopes of -1.59$\pm$0.03 and -4.16$\pm$0.04 for the density-dominated region and velocity-dominated regions, respectively. Similarly, we report slopes of -1.78$\pm$0.02 and -3.98$\pm$0.04 for T2t0. The break points between power-law fits are ${\rm log~}k_v =$ -0.20$\pm$0.02 for W1T1t0.2, and -0.01$\pm$0.01 for T2t0. 
 The shaded region indicates the standard deviation of the data in each bin.
\label{vcsfig}}
\end{center}
\end{figure}

Figures \ref{vcafig} and \ref{vcsfig} show the VCA and VCS results, respectively, for outputs W1T2t0.2 and T2t0. VCA produces similarly sloped power laws for both outputs, but there is a constant horizontal offset, which is similar to that for the SPS. This implies that at all spatial scales output W1T2t0.2, the case with feedback, has more energy than output T2t0. Both VCA curves exhibit some curvature, which suggests they could be better fit by a broken power-law. However, VCA theory predicts a single power-law slope \citep{lazarianp04}, so by convention we fit the curves with a single power-law.   The winds produce a slightly flatter VCA slope.  

Like SPS, the VCA distance is the t-statistic between the slopes.  The similar slopes and curve shapes produce a relatively small distance, and we conclude VCA is not useful for characterizing feedback properties. 


In Figure \ref{vcsfig}, VCS also shows a horizontal offset between the two curves. 
However, we also note a difference in both VCS power-law fits, and, more importantly, the break point between the two fits. Physically, this transition point indicates the scale at which the dispersion of the density fluctuations is equal to the mean density, which is also influenced by the optical depth of the gas  \citep{lazarianp04}.  The location of the break thus depends both on the density distribution and the power spectrum of the underlying turbulence \citep{lazarianp08}.  
For W2T2t0.2, $k_{\rm cr}$/(km$^{-1}$s) $\simeq 0.63$, while for T2t0 $k_{\rm cr}$/(km$^{-1}$s) $\simeq 0.98$, which is statistically significant.  However, the VCS distance metric is proportional to the sum of the t-statistics between the two sets of slopes weighted by the fit error,  and it does not depend on the break-point location. Thus, the distance metric as currently defined may miss a significant difference between the outputs. 

The output without feedback 
appears to have a larger range over which it is dominated by density fluctuations ($k_v$/(km$^{-1}$s) $\lesssim 0.98$). The density-dominated regime is smaller for the run with feedback, such that changes in gas density affect a smaller portion of the structure apparent in the cloud emission. Since the winds inject energy and create additional expansion of $\sim 1-3$ km/s, it makes sense that the velocity-dominated regime extends to smaller $k_v$, which corresponds to a larger effective $V_{\rm ro}$.

Irrespective of the break-point, the velocity-dominated regime should follow a power-law set by the underlying velocity structure function. For supersonic shocks, we expect $P(k_v) \propto k_v^{-4}$ with the slope steepening for $k_v> \Delta V_{r_0}^{-1}$ depending upon the shape of the line profile \citep{lazarianp08}. The fits for both outputs in the velocity dominated regime are consistent with this prediction.  
Indeed, we find that the slopes are relatively similar above and below the break. Thus, variation in the breakpoint location could provide insight into the underlying turbulent driving scale.

\subsubsection{Bispectrum/Bicoherence}

The bispectrum measures both the magnitude and phase correlation between Fourier signals. This gives it a distinct advantage over two-point correlation methods such as VCA and VCS, which do not preserve phase information. Consequently, the bispectrum is useful to quantify nonlinear wave-wave interactions, which may be prevalent in turbulent magnetized gas \citep{burkhart09}. 

The bispectrum is obtained by computing the Fourier transform of the three-point correlation function. In our analysis, we use the bispectrum to calculate the bicoherence, a real-valued, normalized summary. The bicoherence also encapsulates the amount of phase coupling on different scales. Thus, it is a more straightforward metric than the bispectrum for comparing two datasets. Following the analysis of K16, we generate sets of randomly sampled spatial frequencies that are sampled on scales up to half of the image size (i.e., 127 pixels). 
For each output, we compute the bicoherence of the integrated intensity maps using the random sets. 

\begin{figure}[h!]
\begin{center}
\includegraphics[width=1\columnwidth]{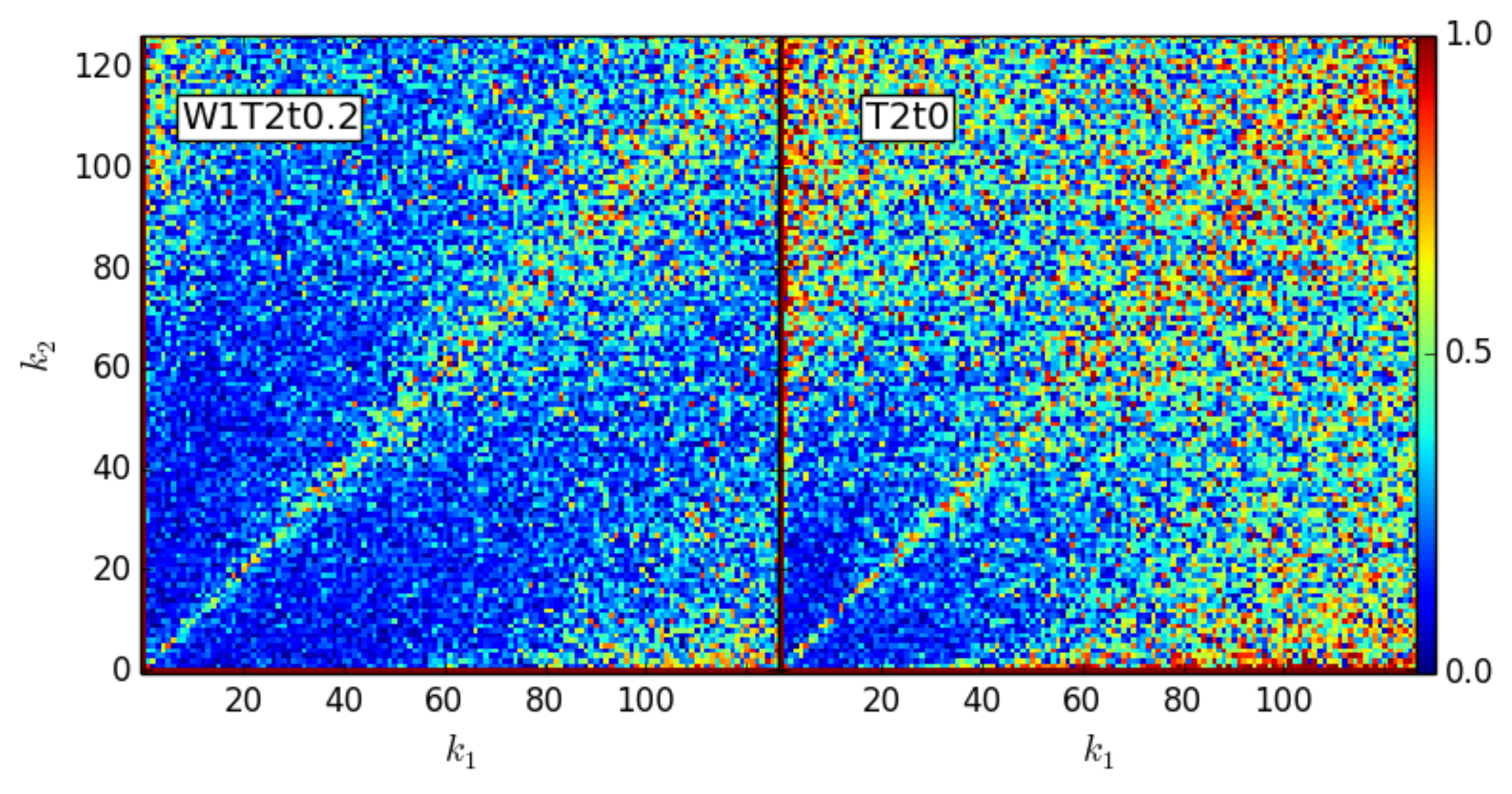}
\caption{\selectlanguage{english} Bicoherence matrices for outputs W1T2t0.2 and T2t0. 
We calculate the bicoherence over 100 randomly sampled spatial frequencies, denoted by $k_1$ and $k_2$.  The colorscale denotes the bicoherence magnitude and the degree of correlation between wavenumbers $k_1$ and $k_2$: 
a value of 0 indicates random phases, i.e., no correlation,  while a value of 1 indicates strong phase coupling.  
\label{biofig}}
\end{center}
\end{figure}

Figure \ref{biofig} depicts the bicoherence matrices for outputs W1T2t0.2 and T2t0. The bicoherence matrix of W1T2t0.2 exhibits a clear signal on the diagonal; this is the trivial case of $k_1=k_2$. However, it exhibits little correlation elsewhere. In contrast,  the bicoherence maxtrix of T2t0 shows enhanced correlation for large wavenumbers (small scales). In general, it contains a significant fraction of pixels above 0.5, which suggests fairly widespread correlation.  If magnetic waves enhance correlation across scales, the wind shells may break up the volume and, thus, reduce correlation. Although shell expansion may perturb the magnetic field and excite magnetosonic waves, it is difficult to see any direct evidence of this against the background of the initial turbulence  (OA15). The comparison of the two bicoherence matrices in fact seems to suggest that the shells reduce correlation perhaps by disrupting the propagation of MHD waves.

The bicoherence distance metric is defined to be a function of the point-by-point differences between the two bicoherence matrices (specifically the $L_2$ norm). Thus, varying structure or degree of correlation in the bicoherence as illustrated in Figure \ref{biofig} increases the distance.

In past work there has been some suggestion that the bispectrum is sensitive to feedback. \citet{burkhart10} computed the bispectrum of HI maps of the SMC. They found that HI column density maps exhibit higher bispectrum amplitudes, which may imply stronger correlations, compared to a turbulent Gaussian random field. They also discovered a break around $\sim 160$ parsecs, where the correlation decreases, a signature  which they attributed to expanding supernovae shells. \citet{burkhart10} also demonstrated that correlation is higher for super-Alf\'venic turbulence ($\mathcal{M}_A =\sqrt{12\pi \rho} \sigma/B> 1$). The Alf\'ven Mach numbers of our outputs range from $\sim 1-5.5$. Since the velocity dispersion, and hence the Alf\'venic Mach number,  increases for the strong feedback case, we would a priori expect {\it more} correlation. However, we see the opposite. This supports the conclusion that the shells suppress the free propagation of MHD waves and reduce scale coupling.



\subsubsection{$\Delta$-Variance}

The $\Delta$-variance is a filtered average over the Fourier power spectrum \citep{stutzki98}. It has been used to characterize the structure distribution and turbulent power spectra of molecular cloud maps. The revised method presented by \citet{ossenkopf08a,ossenkopf08b} takes into account noise variation and provides a means to discriminate between small-scale map structure and noise, so K16 adopt this approach in {\sc turbustat}. To compute the $\Delta$-variance, we generate a series of Mexican hat wavelets that vary in width. We approximate each wavelet as the difference between two Gaussians with a diameter ratio of 1.5 as recommended by \citet{ossenkopf08a}. For each output, we weight the integrated intensity map by its inverse variance, convolve it with a Mexican hat wavelet, and calculate the $\Delta$-variance in Fourier space. 

\begin{figure}[h!]
\begin{center}
\includegraphics[width=1\columnwidth]{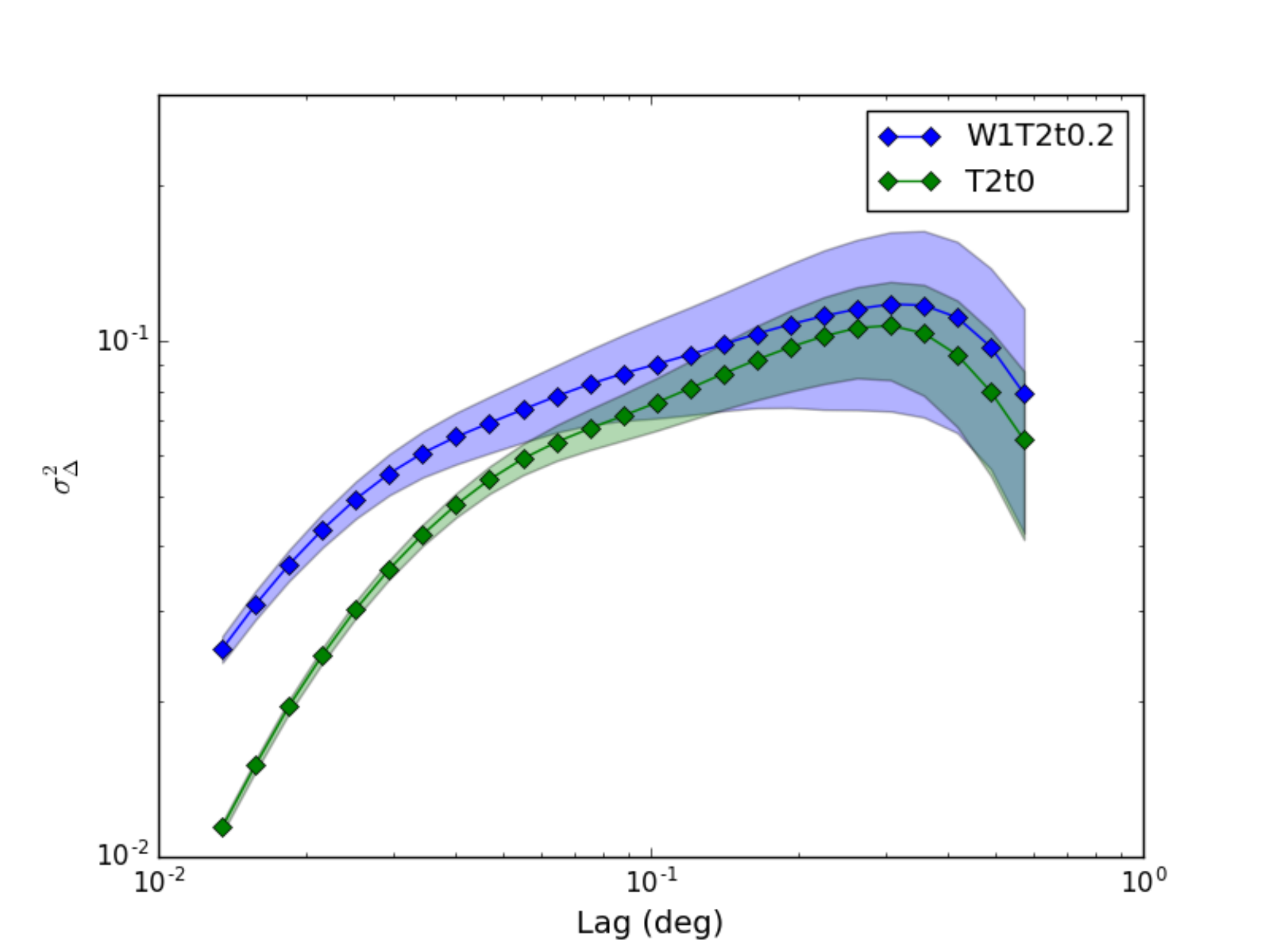}
\caption{\selectlanguage{english}$\Delta$-variance spectra for outputs W1T2t0.2 (blue, top) and T2t0 (green, bottom), where the ``Lag" denotes the width of the Mexican hat wavelet used in the convolution.  The shaded region indicates the standard deviation of the data in each bin.
\label{delvarfig}} 
\end{center}
\end{figure}

Figure \ref{delvarfig} shows the $\Delta$-variance as a function of wavelet width, which is termed the ``lag" by convention \citep{stutzki98}. 
The $\Delta$-variance curve of T2t0 declines more for lags below 0.1 degrees than the curve for output W1T2t0.2, such that the curve shapes are noticeably different.  The $\Delta$-variance distance metric is defined as a function of the total differences between the two curves (the $L_2$ norm). Thus, any offset or change in slope as shown in Figure \ref{delvarfig}  increases the distance.

In noisy observations, the $\Delta$-variance increases towards small lags, indicating enhanced structure. Here, the difference between the curves indicates that the wind output has slightly more structure on smaller scales, a result probably caused by the wind shells, which have a thickness of a few pixels (0.1 pc). However, the $\Delta$-variance of W1T20.2 does not exhibit any break, which would indicate a preferred structure scale. In fact, it more directly resembles a pure power law than the non-wind $\Delta$-variance curve. \citet{ossenkopf08b} also found a smooth power-law $\Delta$-variance for rho Ophiuchus, even though clump-finding on the same map produced a mass distribution with a break \citep{motte98}. These results suggest that the $\Delta-$variance statistic as applied to integrated intensity maps is only relatively sensitive to feedback signatures.

\subsubsection{Wavelet Transform}

Wavelet transforms offer an alternative data decomposition to Fourier transforms for studying intermittency and nonlinear scale coupling. Wavelet transforms have been utilized to study MHD and plasma turbulence for more than two decades \citep{farge15}. They are less frequently applied in studies of astrophysical turbulence, although the first application of the wavelet transform was presented for $^{13}$CO molecular emission of L1551 by \citet{gill90}. Here, we define the wavelet transform as the average value of the positive regions of a convolved image (K16); it is essentially an intensity average computed over a range of size scales. We convolve the integrated intensity maps of the outputs with a Mexican hat kernel, a process similar to that of the $\Delta$-variance technique described in \S3.2.4. 

\begin{figure}[h!]
\begin{center}
\includegraphics[width=1\columnwidth]{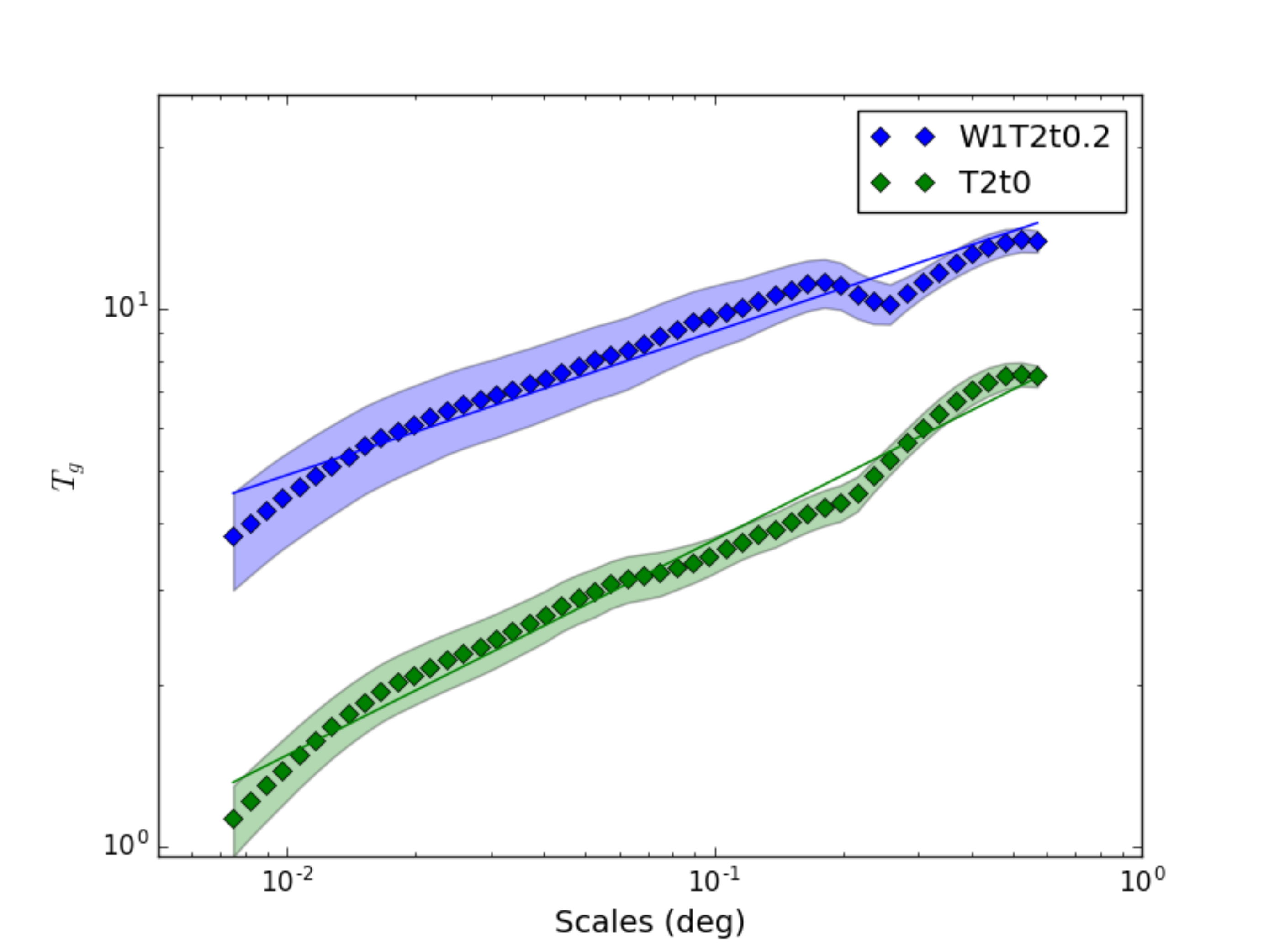}
\caption{\selectlanguage{english}Wavelet transform for outputs W1T2t0.2 (blue) and T2t0 (green) as a function of the Mexican hat wavelet width $a$. The lines indicate the best-fit power-law for the range $-2.03<log(a)<-0.43$.  
We report slopes of 0.267$\pm$0.008 for W1T2t0.2 and 0.400$\pm$0.008 for T2t0.
 The shaded region indicates the standard deviation of the data in each bin.
\label{wavefig}}
\end{center}
\end{figure}

Figure \ref{wavefig} shows the wavelet transform for the fiducial outputs. Following K16, we fit a portion of the transform to a power-law, where the range is informed by the results of \citet{gill90}.  Although the resultant slopes are similar, output T2t0, the purely turbulent model, diverges more from power-law behavior than output W1T2t0.2. We also note that the wavelet transforms are higher for output W1T2t0.2 than than T2t0, which is consistent with stronger molecular excitation resulting from the higher density and temperature in the wind shells. 

The wavelet distance metric definition is identical to that of the SPS, VCA and VCS statistics; the distance is the t-statistic of the difference in the slopes. The variations in curvature shown in Figure \ref{wavefig} increase the distance provided they impact the overall fit, while the offsets will be ignored.

While the shape of the wavelet transform may provide insight into  underlying turbulent properties, neither the offset nor the slope appear to exhibit sufficiently different behavior to serve as a diagnostic for embedded feedback. Indeed, the first astrophysical application of the wavelet transform by \citet{gill90} compared the wavelet transform both ``on" and ``off" the outflow region of L1551; however, they found the slope of the gas associated with the outflow was slightly flatter than the non-outflow gas. We find a similar trend although our slopes are very different - possibly because we analyze $^{12}$CO rather than $^{13}$CO.  Their data exhibited a turnover around $log(a) \sim -0.6$, which they postulated was a transition between two competing physical processes. The turnover here is more subtle, but it occurs at a similar point for both outputs, so it more likely represents edge effects. Given the similarity between the two curves, we tentatively conclude that this formulation of wavelet analysis is not a strong indicator of feedback.


\subsection{Morphology Statistics}

In this section we present statistics quantifying the morphology of the emission distribution, namely, the Genus statistic and dendrograms.

\subsubsection{Genus Statistic}

Genus statistics characterize spatial information in a data map by identifying and counting local minima and maxima. They essentially compute the difference between the number of isolated features (peaks) and holes (voids) above and below a given threshold, respectively. Beginning with the work of \citet{gott86}, Genus statistics have been frequently used in cosmological studies to characterize the distribution of mass in the universe. \citet{kowal07} were the first to apply them to interstellar turbulence. They analyzed density and column density maps produced by MHD simulations and found that the shape of the distribution correlates with the sonic Mach number. This analysis was extended to observational data of the Small Magellenic Cloud (SMC) by \citet{chepurnov08}, who noted the statistic could be sensitive to the presence of shells.

To compute the Genus statistic for each output, we normalize the integrated intensity map and convolve it with a 2D Gaussian kernel with width of 1 pixel. This smooths the map so that small scale variations and noise do not contribute to the number of features.
We then divide the intensity range $[I_{\rm min}, I_{\rm max}]$ into 100 evenly spaced threshold values and compute the Genus for those values above 20 percent of the minimum intensity. We fit the distribution with cubic splines of equal bin size for intensities $<4$ K $\kms$, which is the maximum threshold for the purely turbulent case.

\begin{figure}[h!]
\begin{center}
\includegraphics[width=1\columnwidth]{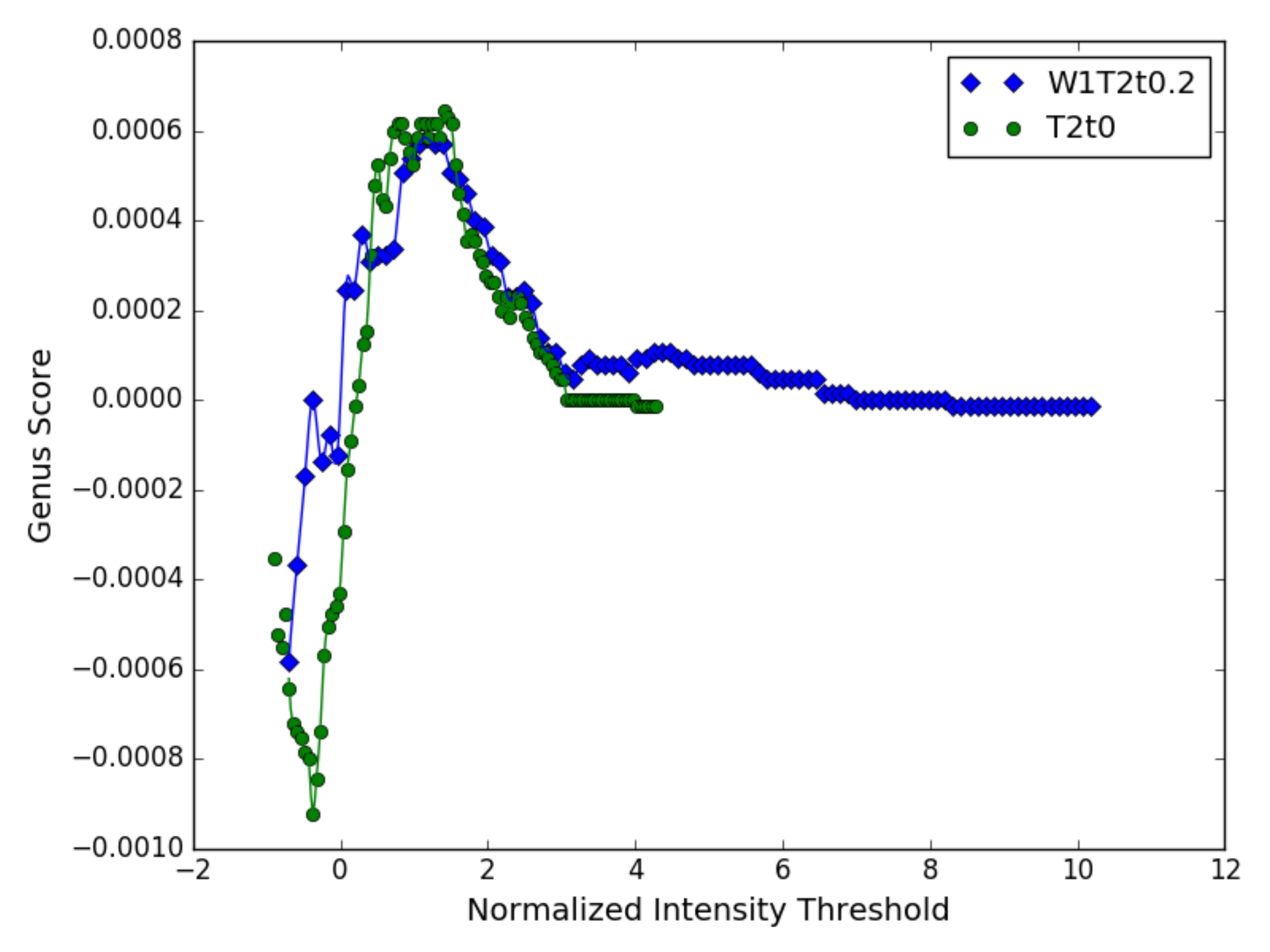}
\caption{\selectlanguage{english} Genus score for outputs W1T2t0.2 (blue diamonds) and T2t0 (green circles). The data are normalized such that the $x$-axis indicates the number of standard deviations from the mean  and the score is normalized by the number of pixels in the integrated intensity map.
\label{genusfig}}
\end{center}
\end{figure}

Figure \ref{genusfig} shows the Genus as a function of intensity threshold for the two fiducial outputs. Positive values indicate a relative excess of peaks (a ``clump-dominated" topology), while a negative Genus indicates an excess of voids. We find both curves exhibit similar behavior for normalized intensities below 4. As expected, output W1T2t0.2 has a broader range of intensity values due to the higher velocities and excitation in the wind shells, and thus, exhibits some structure for higher integrated intensities. Between thresholds of -1 and 1, the Genus is smaller for the purely turbulent model, which indicates that there are more voids in the  emission compared to the case with feedback. The Genus for W1T2t0.2 is higher at low-intensities, but this may be because the voids created by winds are larger than those created by pure turbulence, such that the total number of minima is reduced. This effect would likely be enhanced for real clouds, where winds can break out and create sight-lines nearly empty of molecular emission (A11). 

K16 define the Genus distance metric as the average of the absolute value of the differences between the Genus curves. The larger the disparity in the number of peaks and voids between two datasets, the larger the distance. Thus, the discrepancy between the two curves at low intensities shown in Figure \ref{genusfig} increases the distance between the outputs. 

Our analysis highlights one advantage of the Genus statistic: it is sensitive to both over-densities and voids.  \citet{chepurnov08} analyzed HI data of the SMC, which visually displays a large number of expanding shells with sizes of $\sim 100$ pc. The shells were not apparent at small scales ($< 100$ pc), but at intermediate scales (120-200 pc) the Genus had a neutral or slightly positive value, which they attributed to shells.  
In practice, however, the thickness and morphology of clouds varies significantly between different star forming regions. In our comparison, the difference between the two curves is relatively subtle. Thus, it may be most informative when employed to compare sub-regions within clouds.


\subsubsection{Dendrograms}

Dendrograms are hierarchical structure trees, which may be created for both 2D and 3D data \citep{rosolowsky08}. Here, we compute dendrograms of the 3D spectral cubes. The number of peaks (``leaves") and the number of hierarchical levels (``branches") are useful metrics that prior studies have shown to be sensitive to underlying physics, including gravity and magnetic field strength, as well as emission properties \citep{goodman09,burkhart13b,beaumont13}.
Here, we use dendrograms to characterize the hierarchical structure of the emission. 
To create the dendrogram, we first identify the peak intensity value in the data and then proceed to smaller intensities and successively catalog local maxima. The leaves on the same level of hierarchy are connected by a branch. 
To account for simulated noise in our data maps, we set a minimum distance between two local maxima, $\delta_{\rm min}$. Increasing $\delta_{\rm min}$ decreases the total number of features, i.e., it ``prunes" the tree \citep[e.g.,][]{burkhart13b}. 

K16 consider two dendrogram statistics: the number of features or leaves and the histogram of leaf intensities. To compute the first statistic, we generate multiple dendrograms per output by varying $\delta_{\rm min}$ from $10^{-2.5}$ K$-$10$^{2}$ K in 150 logarithmic steps. We then count the total number of leaves associated with each $\delta_{\rm min}$.  To compute the second statistic, we create a series of dendrograms for the outputs using the same range of $\delta$, but instead of counting features we produce histograms of the leaf intensities for each value. 
We renormalize the intensities so that the mean of the histogram occurs at zero.

\begin{figure}[h!]
\begin{center}
\includegraphics[width=1\columnwidth]{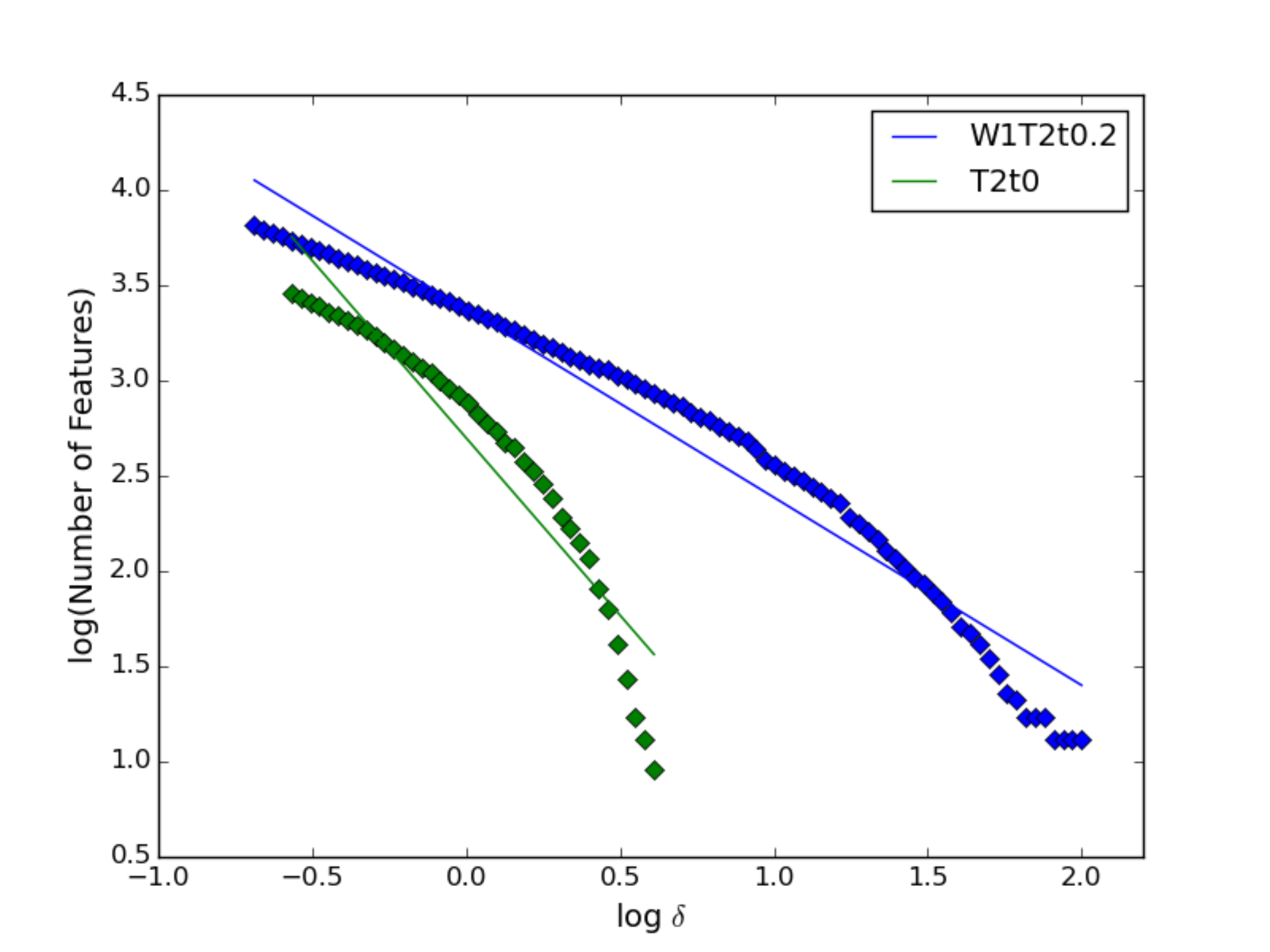}
\caption{Number of dendrogram features as a function of intensity spacing, $\delta$, for output W1T2t0.2 (top, blue) and T2t0 (bottom, green). The lines indicate power-law fits to each curve, having slopes -0.99$\pm$0.02 and -1.9$\pm$0.1 for runs W1T2t0.2 and T2t0, respectively. 
\label{dendro1}}
\end{center}
\end{figure}

Figure \ref{dendro1} displays the number of leaves as a function of $\delta$ for the two fiducial outputs.  Output W2T1t0.2 follows a power-law up to $\delta \sim 10$, while output T2t0 deviates from a power-law at $\delta_{\rm min}\sim$1 K. The latter trend agrees with the results of \citet{burkhart13b}, who analyzed MHD turbulence simulations and found that the number of leaves significantly declines as $\delta$ increases. They also demonstrated that the power-law index for larger $\delta$ values steepens from -1.1 to -3.9 as the sonic Mach number declines from 7 to 0.7. The shape of the curve for output W2T1t0.2 suggests an interesting signature of winds; feedback increases the number of leaves significantly for large $\delta$.  As a result, we also find that the output with feedback contains more structure than the purely turbulent output at all scales. 

The distance metric for the number of features is defined following the convention of the SPS and other power-law statistics, where the distance is the t-statistic of the difference in the slopes. This means the disparate slopes illustrated by Figure \ref{dendro1} produce a large distance. However, the distance is not directly sensitive to the details of the curve shape, which is codified by the power-law fit. Consequently, dissimilarities in the intensity range, which reflect the amount of compression, do not contribute to the distance.  

\begin{figure}[h!]
\begin{center}
\includegraphics[width=1\columnwidth]{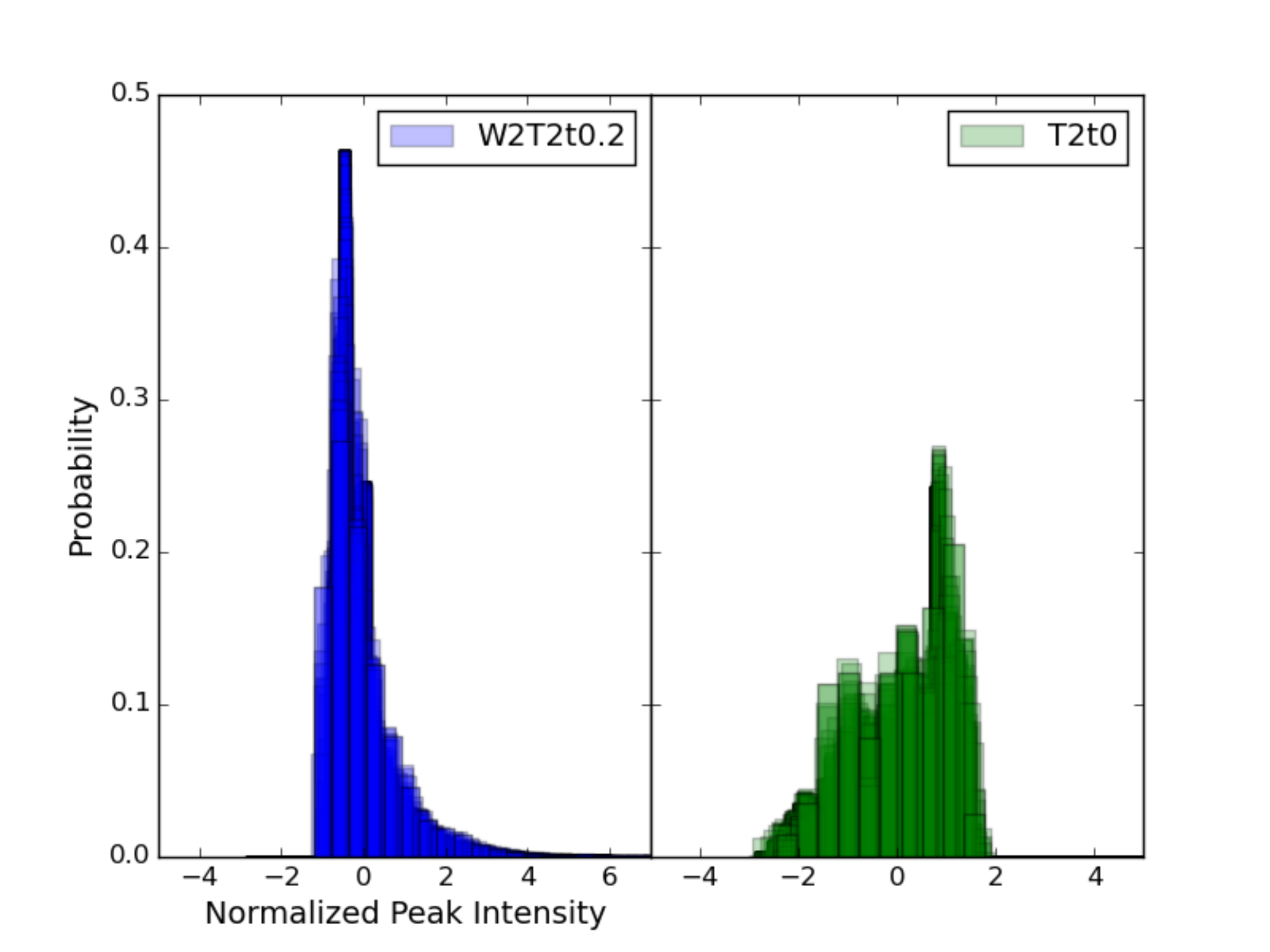}
\caption{Histograms of the renormalized dendrogram peak leaf intensities. The distributions for all $\delta_{\rm min}$ values are stacked.
\label{dendro2}}
\end{center}
\end{figure}

The distribution of peak intensities provides additional insight into the emission structure.
Figure \ref{dendro2} shows superimposed histograms of the two fiducial outputs for all $\delta$. The two outputs yield significantly different distributions. The histograms of T2t0 contain a wider range of leaf values than those of  W1T2t0.2, whose histograms are all strongly peaked on the mean value. Output W1T2t0.2 also produces a long tail of high intensity values. 

The distance metric depends on all bin-wise differences and is defined as the Hellinger distance for each pair of histograms with a given $\delta$ averaged over all values of $\delta$. Thus, large differences between two histograms with a particular $\delta$ will influence the distance. However, the largest distances will be produced by systematic differences between histogram sets, such as those illustrated in Figure \ref{dendro2}. Like the previous dendrogram statistic, the distribution indicates that feedback systematically increases the amount of hierarchy in the emission, which may serve as a signpost of feedback.

\section{Statistical Analysis}\label{distance}

We use pseudo-distance metrics, which were briefly described in the previous section, to efficiently study differences between all synthetic observations. As stated in \S\ref{toolkit}, a pseudo-distance is essentially a single value that encapsulates the degree of difference between two metrics, and it can be used to quantitatively compare the outputs of various spectral cubes. In \S\ref{comparisons}, we identified qualitative differences between a simulation with strong feedback and one with no feedback  and discussed how the distance metric may reflect these changes. We now expand upon this to quantify all simulation differences and determine the sensitivity of the statistics to stellar mass-loss rate, magnetic field strength and evolutionary time. This allows us to check if the previously identified features actually pinpoint signatures of feedback or instead correspond to other combinations of simulation parameters.

For each statistic, we produce a color-plot showing the distances between all simulation pairs (see Figures  \ref{color_inten},  \ref{color_fourier} and  \ref{color_morph}). The color-plots can be interpreted as follows. Each colored square represents the distance between one simulation pair, denoted by the horizontal and vertical indices. The colorbar indicates the distance values, whose range depends on the statistic as defined in K16.  We arrange the simulations in order to easily compare strong wind models (W1) with weaker wind models (W2) or purely-turbulent models (TXt0). 

Since our limited parameter sampling doesn't allow a rigorous analysis of effects (as in \citealt{yeremi14}), we perform a qualitative assessment of the tools.  We find the importance and detectability of feedback produces a clear signature that would persist in a full parameter space study. Time evolution and magnetic field strength produce weaker signals in the pseudo-distance results, and consequently, their impact is less clear. Table \ref{stats_results} provides a summary of our findings, which we discuss in \S\ref{inten_stat}, \S\ref{fourier_stat}, and \S\ref{morph_stat}.

\begin{deluxetable*}{ c|c|ccc }
\tablecaption{ Statistic Sensitivity\tablenotemark{a} \label{stats_results}}
\tablehead{ \colhead{Family} & \colhead{Statistic} & \colhead{Wind Activity} & \colhead{Magnetic Field}  & \colhead{Time Evolution}}
\startdata
 		& Probability Distribution Function (PDF) &  $\checkmark$   & $\cdots$  & $\cdots$ \\
		& PDF Skewness					& $\checkmark$ & $\cdots$ & $\sim$  \\
Intensity 	& PDF Kurtosis 					&   $\checkmark$ & $\cdots$ & $\sim$  \\
Statistics  	& Principal Component Analysis (PCA)	&  $\checkmark$ & $\cdots$ & $\sim$  \\
		 & Spectral Correlation Function (SCF) 	& $\checkmark$ & $\cdots$ & $\sim$   \\
		 & Cramer & $\checkmark$ & $\checkmark$  & $\cdots$    \\ \hline
  		 &  Spatial Power Spectrum (SPS) & $\sim$ & $\cdots$ &  $\checkmark$  \\
     		 &  Velocity Channel Analysis (VCA) &  $\sim$ & $\cdots$ & $\sim$ \\ 
Fourier    	& Velocity Coordinate Spectrum (VCS) & $\sim$ & $\sim$ & $\cdots$  \\ 
Statistics 	& Bicoherence &  $\checkmark$ &   $\cdots$ & $\sim$     \\ 
		& $\Delta$-Variance & $\sim$  & $\cdots$ &  $\sim$   \\
		& Wavelet-Transfrom &   $\checkmark$ & $\sim$ & $\sim$ \\  \hline
		&  Genus & $\cdots$ & $\sim$ & $\sim$ \\  
Morphology& Dendrogram Leaves & $\checkmark$ & $\cdots$ & $\sim$  \\  
Statistics	 & Dendrogram Feature Number & $\sim$ & $\cdots$ & $\sim$
\enddata
\tablenotetext{a}{A summary of the statistic responses to the three primary physical effects. We characterize the response as strong ($\checkmark$), weak ($\sim$) or unclear ($\cdots$). A statistic is strong if it shows a clear monotonic trend, a statistic is weak if it shows a slight response, a statistic is unclear if a statistic is appears to show no sensitivity or if the trend is non-monotonic.}
\end{deluxetable*}

\subsection{Intensity Statistics}\label{inten_stat}

\begin{figure*}[h!]
\begin{center}
\includegraphics[width=0.95\columnwidth]{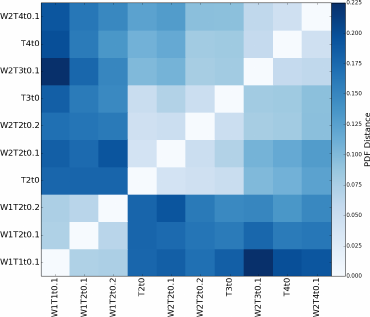} \hspace{0.1in}
\includegraphics[width=0.95\columnwidth]{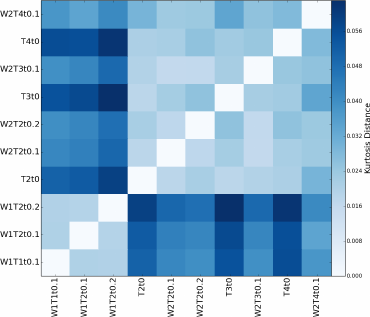} 
\includegraphics[width=0.95\columnwidth]{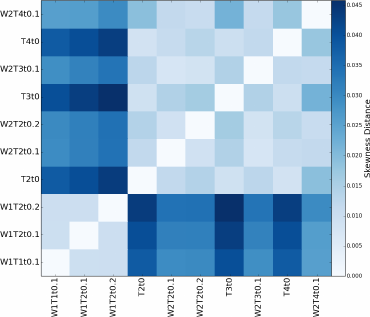} \hspace{0.1in} 
\includegraphics[width=0.95\columnwidth]{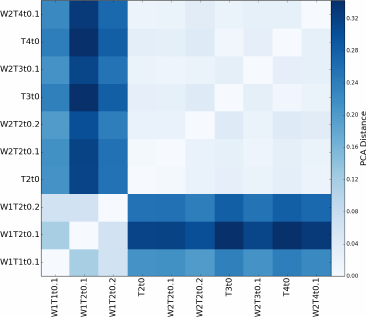}  
\includegraphics[width=1.0\columnwidth]{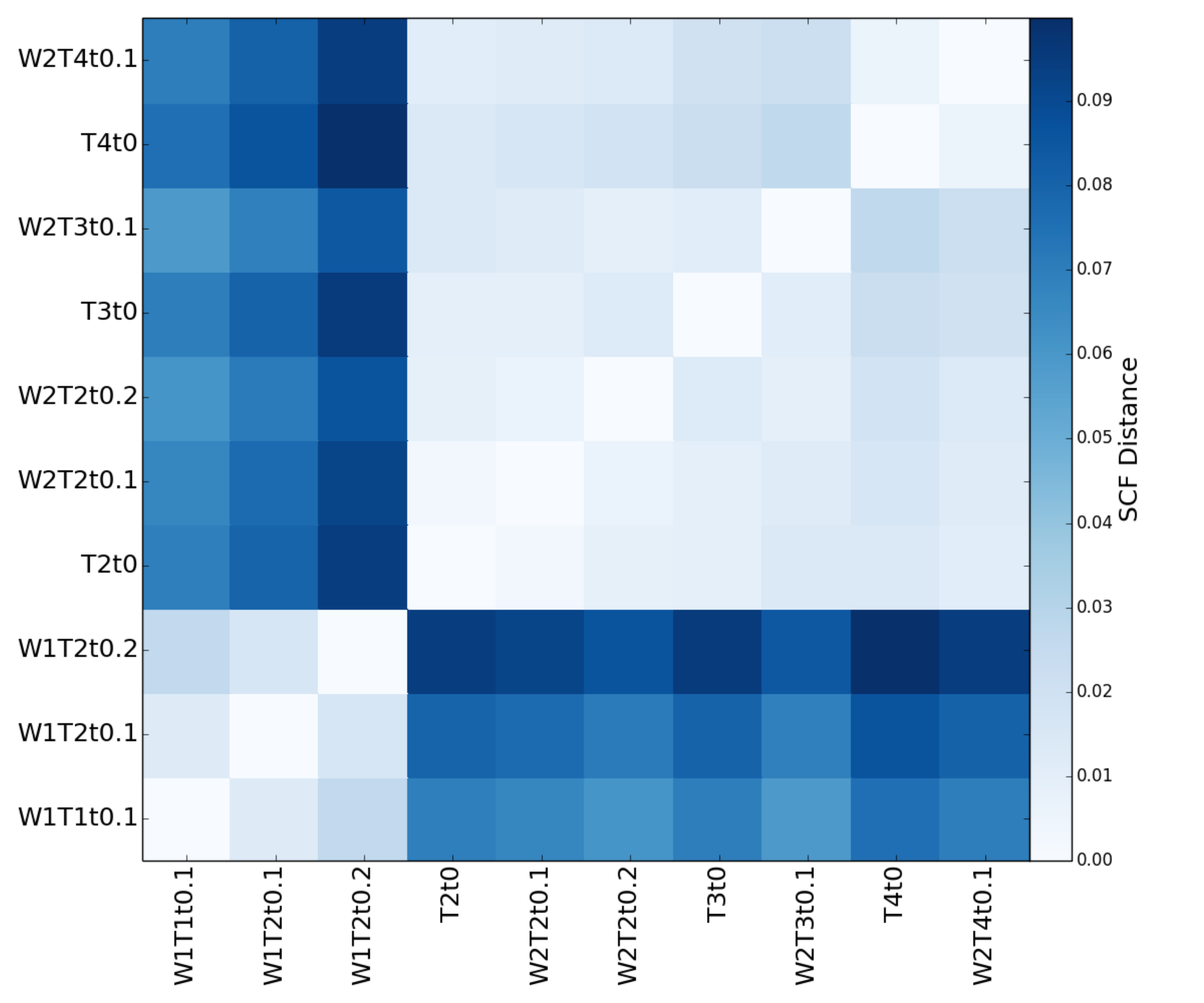} \hspace{0.1in} 
\includegraphics[width=0.95\columnwidth]{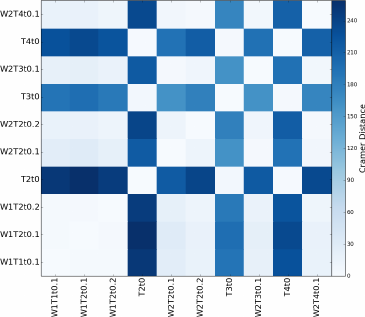} 
\caption{Intensity statistic color-plots. Each statistic utilizes a different distance metric to quantify the difference between two simulations. The colored squares represent the distance between the simulations indicated by the horizontal and vertical indices. \label{color_inten} }
\end{center}
\end{figure*}

We show the color-plots for all intensity statistics in Figure \ref{color_inten}. With the exception of the Cramer statistic, we find that these statistics exhibit strong responses to changes in stellar mass-loss rate. The color-plots show the largest distances appear between any strong wind model (W1) and any weak wind model (W2) or purely turbulent model (TXt0). The kurtosis, skewness, and SCF are the clearest examples of this, as they display a trend among pairs. These statistics yield the largest distances between pairs W1 and TXt0, followed by pairs  W1 and W2. Furthermore, they capture a similar, weaker trend for pairs W2 and TXt0. 


A distance trend with wind strength is less clear for the PDF and PCA statistics. We find that time evolution randomly impacts the magnitude of these statistics' strong wind distances. Weaker wind PDF pairs appear to correlate slightly with magnetic field strength. This does not occur for the PCA and SCF statistics, since their distances between W2 pairs are quite small. Insensitivity to magnetic field strength is consistent with the conclusions of \citet{yeremi14}.


The Cramer statistic is by definition a distance metric, so we include its discussion and analysis here rather than in \S\ref{comparisons_inten}. As described in \citet{yeremi14}, the Cramer  statistic compares the inter-point differences between two data sets with the  differences between points within each individual data set. Following K16, we compute the Cramer statistic using only the top 20\% of the integrated-intensity values. We find that the statistic exhibits a behavior different from that of the other intensity statistics. As Figure \ref{color_inten} shows, the Cramer statistic displays very large distances between purely turbulent outputs and outputs with any degree of feedback. The wind strength appears less important to the statistic than wind presence does, which indicates a binary sensitivity to stellar-mass loss rate. The Cramer statistic is also slightly sensitive to magnetic field strength, as illustrated by the varying distances between the purely turbulent models.

Considering the various degrees of response, we find PCA to be a strong candidate for constraining feedback signatures. As Figure \ref{pca1} shows, this statistic displays sharp, distinct features for a strong wind model, and its color-plot predominantly shows sensitivity to changes in stellar-mass loss rate. The other intensity statistics either exhibit less-distinct features or react to multiple physical changes. Because of this, we recommend using these statistics in concert with PCA. Of the remaining intensity statistics, the SCF is the second most promising candidate, as its color-plot behaves similarly to that of PCA.

\subsection{Fourier statistics}\label{fourier_stat}

Figure \ref{color_fourier} shows all Fourier statistic color-plots. Unlike the intensity statistics, the Fourier statistics do not share a common behavior, and their color-plots appear more heterogeneous. As a whole, we note a variety of sensitivities to changes in stellar mass-loss rate, magnetic field strength, and evolutionary time. The 
wavelet transform color-plot closely resembles those of the intensity statistics, as their greatest responses correspond to changes in stellar-mass loss rates. The $\Delta$-variance shows some discriminating power to the presence of winds, but it is only weakly sensitive to changes in other underlying properties.


The VCS statistic demonstrates roughly equal, weak sensitivity to both stellar mass-loss rate and magnetic field strength.  However, as noted in \S\ref{VCA}, the VCS distance is independent of the fit break-point, which may respond to the presence of feedback. Thus, the color-plot reflects the minimum degree of distance between the outputs.   As its color-plot shows, distances solely quantifying changes in stellar mass-loss rate tend to resemble those explicitly comparing changes in magnetic field. In fact, some of the largest distances involve T4, the run with the strongest magnetic field. We also note large distances between the turbulent clouds T1 and T2 in the presence of strong winds. These clouds have the same magnetic field strength, indicating that the VCS  distance, which is driven by changes in the broken-power law slopes, is generically sensitive to the turbulent conditions. 

We find the SPS to be sensitive to all simulation parameters, but unlike the VCS, it's responses are not monotonic. Thus, it does not serve as a good diagnostic of feedback.

We find the bicoherence statistic to exhibit a strong response to changes in stellar mass-loss rate, magnetic field strength, and evolutionary time. As time evolves, the wind models do become more alike, and they remain distinct from outputs without feedback. The turbulent models also appear relatively similar to each other. 


Of all of the Fourier statistics, the VCA statistic demonstrates the weakest sensitivity to magnetic field strength. Its color-plot appears insensitive to turbulent structure, as distances only change with wind model and evolution time. 
The color-plot shows the distances for strong wind models to be different from those of all other models. However, as time evolves, the weak wind distances more closely resemble the strong wind distances. This trend is clear because of the magnetic field's weak impact on the distances.  

In summary, despite various degrees of response, many of the Fourier statistics fail to produce distinct signatures corresponding to feedback. As discussed in \S3.2, the most common difference manifests as a horizontal offset, which is a relatively minor change and may naturally occur between two observed clouds.  The exception is VCS, where variation in the pseudo-distance metric correlates with changes in the location of a power-law break.

\begin{figure*}[h!]
\begin{center}
\includegraphics[width=1\columnwidth]{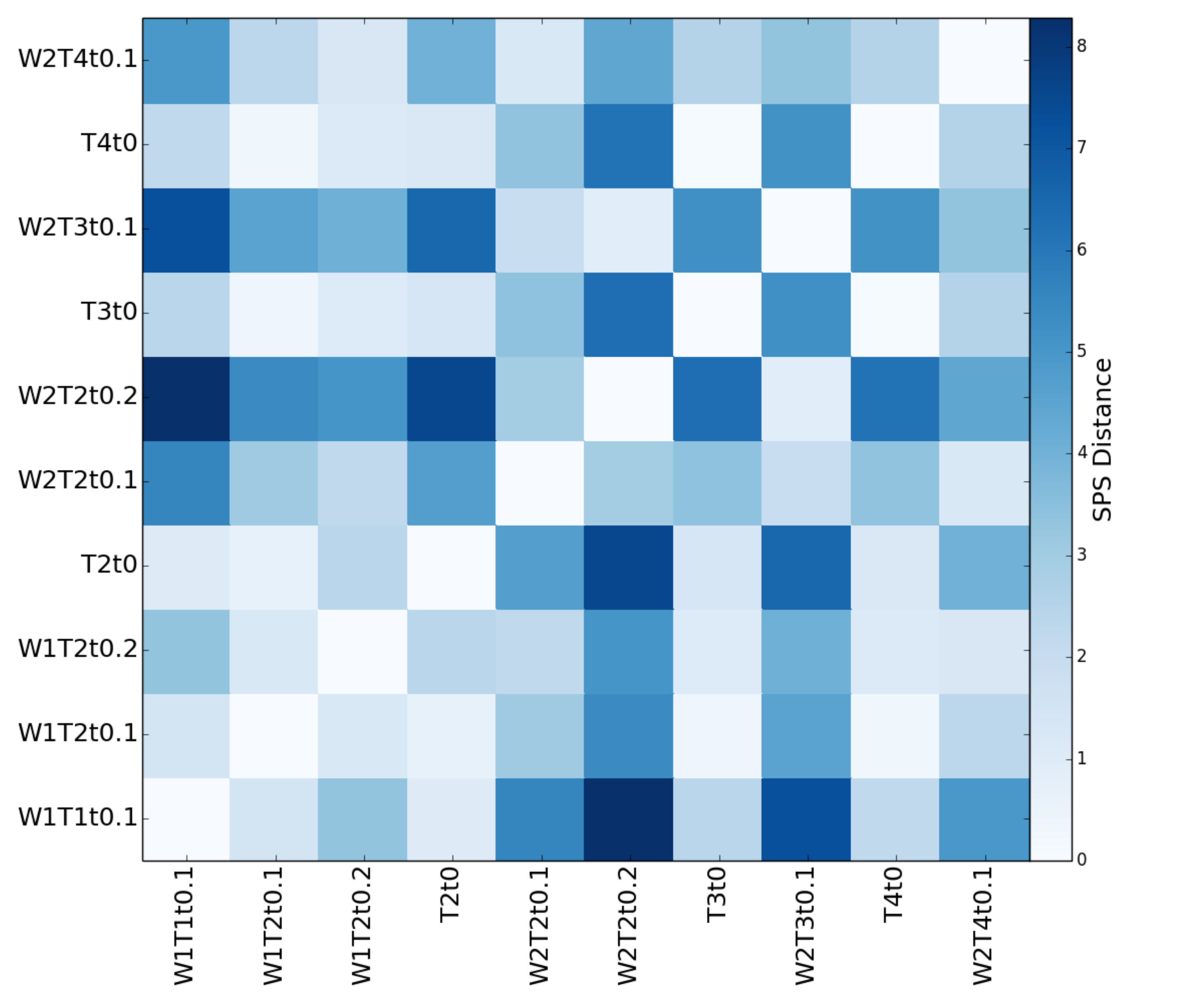}
\includegraphics[width=0.95\columnwidth]{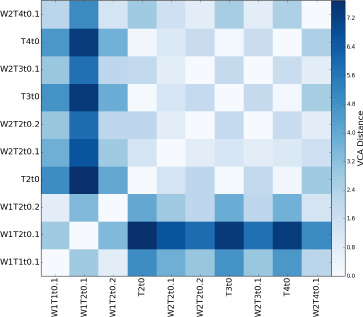} \hspace{0.1in} 
\includegraphics[width=0.95\columnwidth]{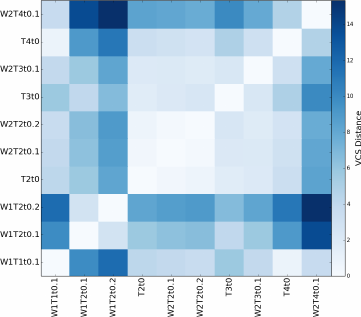}
\includegraphics[width=0.95\columnwidth]{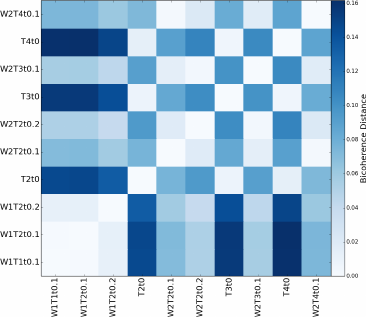} \hspace{0.1in} 
\includegraphics[width=0.95\columnwidth]{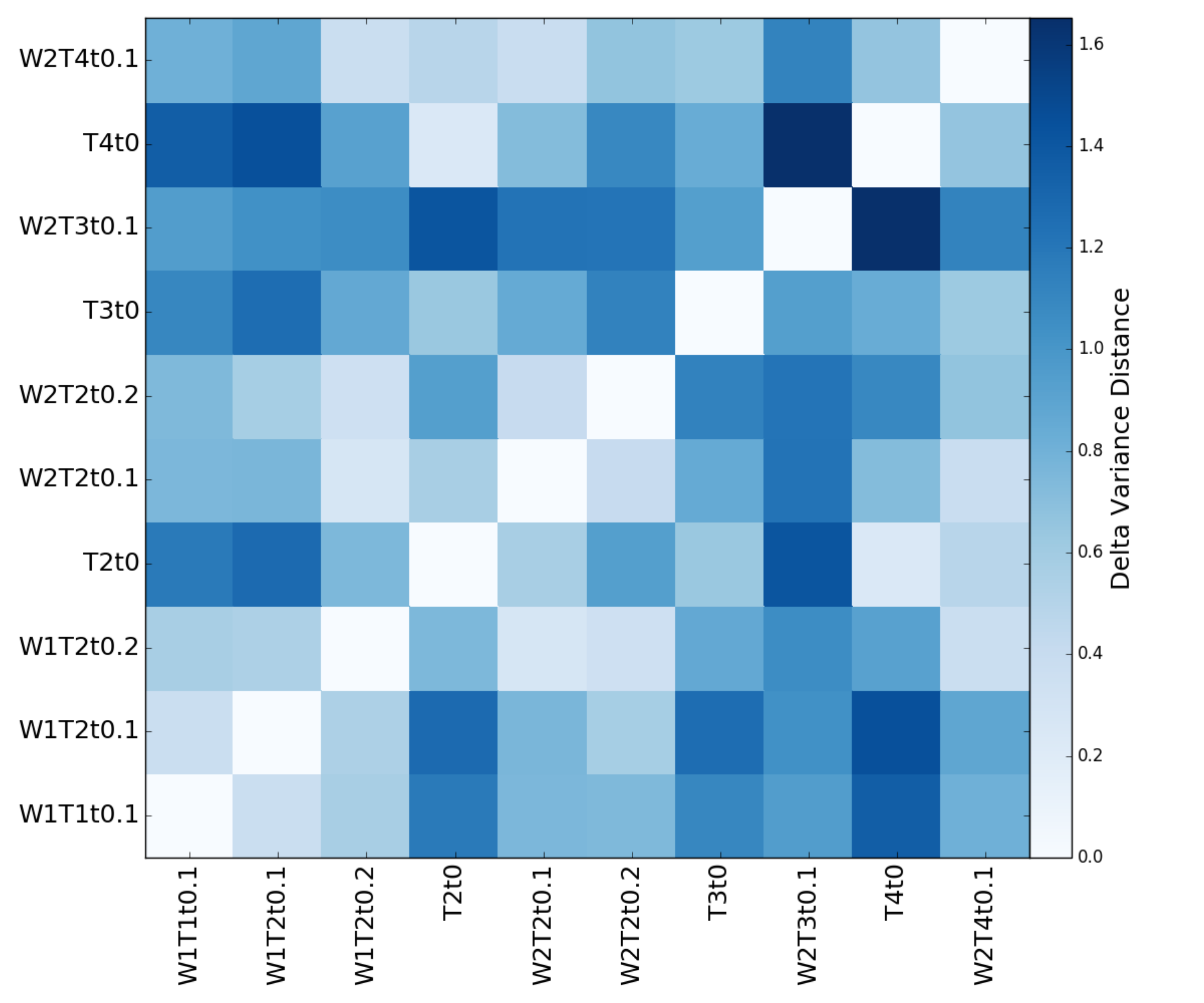}
\includegraphics[width=0.95\columnwidth]{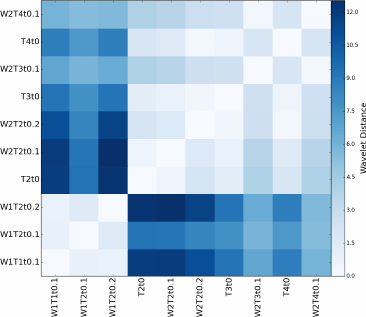} \hspace{0.1in} 
\caption{Fourier statistic color-plots.\label{color_fourier.pdf}%
}
\end{center}
\end{figure*}

\subsection{Morphology statistics}\label{morph_stat}

\begin{figure}[h!]
\begin{center}
\vspace{0.3in} 
\includegraphics[width=0.95\columnwidth]{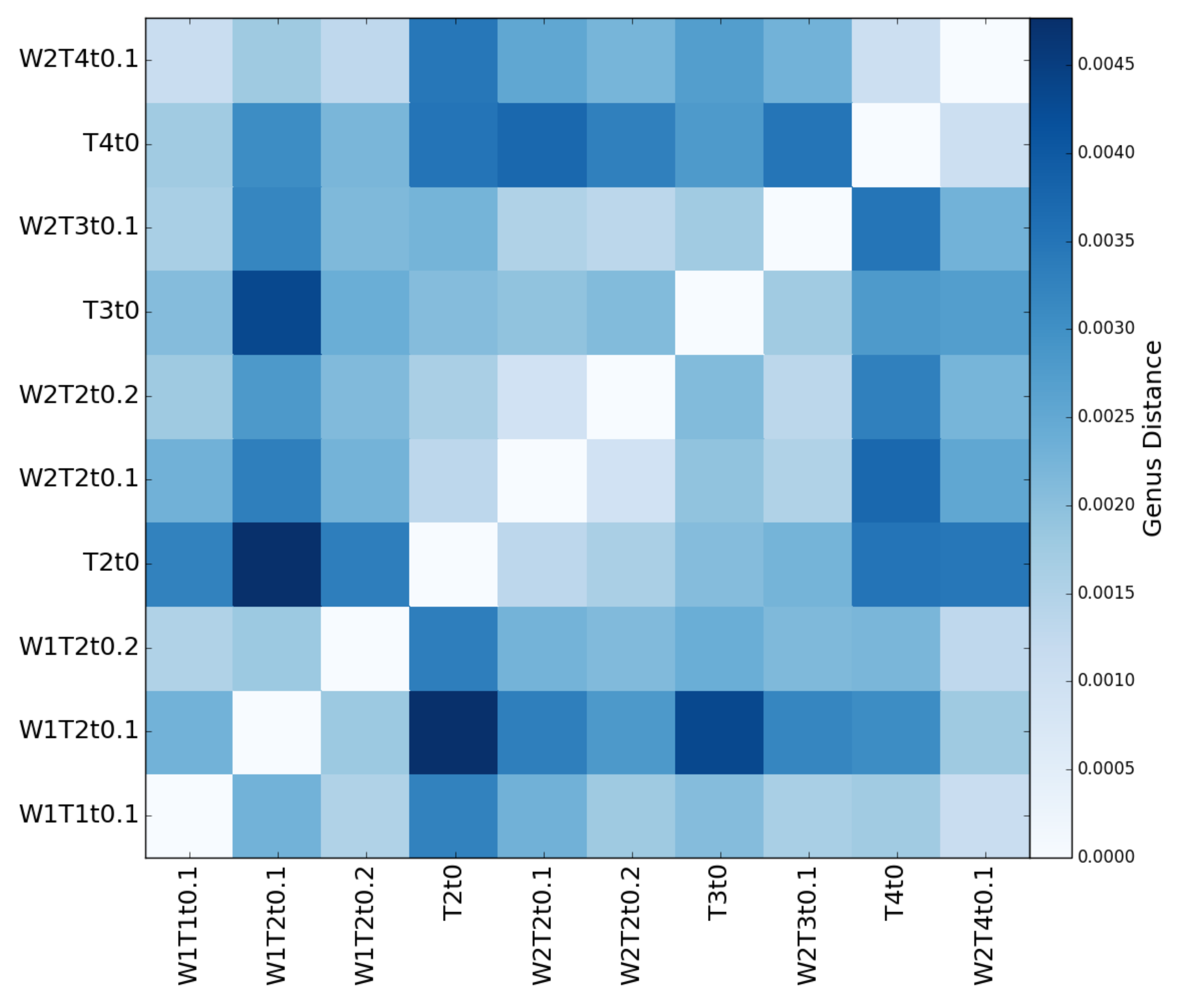}  
\vspace{0.1in} 
\includegraphics[width=0.95\columnwidth]{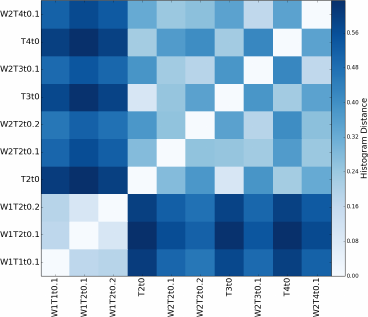} 
\vspace{0.1in} 
\includegraphics[width=0.95\columnwidth]{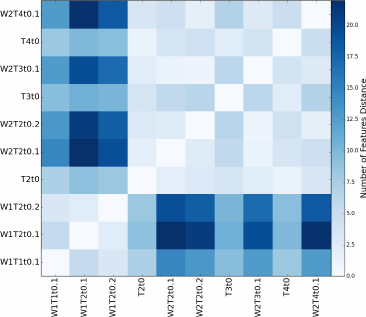} 
\caption{Morphology statistic color-plots.  \label{color_morph}%
}
\end{center}
\end{figure}

Figure  \ref{color_morph} shows the color-plots for the morphology statistics.  Although the Genus statistic produces a wide range of distances, we find that it fails to display any clear trends. Both dendrogram statistics show clear responses to changes in stellar mass-loss rate. The histogram statistic yields the largest distances for strong wind and purely turbulent pairings, followed by strong wind and weak wind pairings. This behavior is similar to that of many other statistics, but the histogram statistic's trend continues within weak wind model pairings, indicating a very clear sensitivity towards wind activity.

The number of features statistic is also sensitive to winds, but it shows opposite correlations between strong wind model pairings. By a significant amount, the largest distances occur for strong wind and weak wind pairings, as opposed to pairings of strong wind and purely turbulent models. A trend does not occur for weaker wind model comparisons, as distances for weak winds and purely turbulent pairs are larger than for pairings between weaker wind models.

Although the histogram statistic produces cleaner trends, we conclude both dendrogram statistics effectively highlight feedback signatures. Both are most sensitive to changes in wind strength, meaning their distances exhibit distinct signatures corresponding to feedback. By comparison, the Genus statistic performs poorly in our formulation. 

\section{Discussion}\label{discussion}

 While this work, together with K16 and \citet{yeremi14}, has made significant strides in consolidating a wide variety of statistics from the literature and systematically testing these statistics across a broad change of physical conditions, several other major dimensions remain unexplored. We discuss these here.

\subsection{Astrochemistry}

 In this study, we focus only on the statistics of global cloud structure as reflected by $^{12}$CO. Other common tracers, including HI, [CI], HCN, N$_2$H$^+$, and other line transitions will produce different statistical trends \citep[e.g.,][]{burkhart10,gaches15}, possibly with unique signatures of stellar feedback. More broadly, the information contained in alternative tracers may help to bridge the gap between atomic and dense molecular gas, and, thus, constrain the influence of various types of feedback from kpc to sub-pc scales. The statistical results for other tracers should be explored in future work.

Our study is also simplified by assuming an extremely basic CO abundance model. The statistics may change with full chemical modeling, which affects the abundance and temperature distributions \citep{glover12,beaumont13,Offner14b}. The consideration of astrochemical networks also introduces an additional set of parameters, including the strength of the UV radiation field and metallicity, which drive changes in the emission \citep{glover12,clark15,bertram15c}. These parameters are likely to impact the statistics and may create degeneracies, for example, in the comparison between high-mass and low-mass star-forming regions. However, as in the case of feedback, only a few statistics have been explored in combination with astrochemical networks. These include PCA \citep{bertram14}, SCF \citep{gaches15}, the centroid velocity structure function \citep{bertram15a} 
and $\Delta$-variance \citep{bertram15b}. Future work is necessary to systematically determine the impact of astrochemical effects on each of the {\sc TurbuStat} statistics.

\subsection{Optical Depth}

The mean density of the simulation and, hence, the corresponding optical depth also influences the shape of the statistics. Optical depth effects have previously been explored and discussed for VCA \citep{lazarianp04,burkhart13a}, velocity centroids \citep{bertram15a}, $\Delta$-variance \citep{bertram15b}, PDFs \citep{burkhart13c} and SPS \citep{burkhart13a}. In general, tracers with lower optical depth, such as $^{13}$CO, reduce projection effects and confusion of cloud structures \citep{beaumont13}. For PDF statistics, high optical depths limit the range of integrated intensities and, thus, lower the distribution width \citep{burkhart13c}. For Fourier statistics, high opacity flattens the spectrum until some saturation point  \citep{lazarianp04,burkhart13a,bertram15a}.  Consequently, we expect significant optical depths to obscure feedback signatures and increase the difficulty of identifying their imprint on the emission. However, the optical depth can be estimated using multiple line transitions or tracers in combination. In instances of high-optical depth, optically thiner, bulk cloud tracers such as $^{13}$CO and [CI] will help to disambiguate feedback and optical depth degeneracies. 

To isolate the impact of optical depth and excitation conditions on the statistics, we also perform the analysis in \S\ref{comparisons} on raw PPV cubes. These results are presented in the Appendix. We find that radiative transfer provides an overall ``stretch" to the data that enhances emission arising from shells created by the winds. In contrast, without the radiative transfer, the distance metrics are relatively similar and features associated with winds from \S\ref{comparisons} (e.g., in the PCA covariance) are absent. 

From an observational perspective these results are unsurprising, as CO has historically been used to study and identify feedback signatures. CO is excited at intermediate ($10^2-10^4$ cm$^{-3}$) densities and becomes optically thick or freezes out at high densities ($\gtrsim 10^4$cm$^{-3}$), coincidentally selecting the subset of molecular gas most impacted by feedback. However, different cloud conditions or density weightings might show some statistical differences without radiative transfer. Future studies are necessary to systematically study the impact of optical depth on the {\sc TurbuStat} statistics.


\subsection{Resolution}

A number of the statistics, especially those sensitive to small scale velocity and density structure, will be sensitive to the simulation resolution. The inertial range of the underlying velocity power spectrum will be larger for higher resolution basegrids \citep[e.g.,][]{kritsuk07}.  K16 compare the {\sc Turbustat} metrics for basegrids of $128^3$ and $256^3$ and find that Cramer, kurtosis, skewness, SCF, SPS, VCA and VCS are sensitive to the change in resolution, differences largely driven by modification to the inertial range. Statistically, the degree of sensitivity is set by the range of $k$ included in the fitting.  Metrics described by power-law functions are fit from several pixels up to half the box size, which reduces the impact of shot noise on small scales and edge-effects on large scales. Within this range, changes in the power-law slope due to either resolution or physics produce a difference.  While the simulations are not solely described by the turbulent power spectrum, those statistics expected to be most sensitive to small scale turbulent structure do register a difference for different resolutions. 

Our study analyzes data from the $256^3$ simulation basegrid, which produces a similar power spectrum to $512^3$ data including the level 1 AMR information (OA15 Appendix). While the extent of the inertial range is important for modeling pure turbulence, our aim here is to identify statistics that can distinguish feedback-related changes between simulations. In the case with winds, OA15 show the dissipation region of the power spectrum is swamped by energy input from the winds on small scales, such that the slope and behavior out to $k\sim 100$ are radically different. This is a sharp signal that accordingly impacts kurtosis, skewness, SCF, SPS, VCA and VCS: those statistics that are demonstrably sensitive to small scale turbulent fluctuations. While we expect simulations with larger basegrids to better model the underlying turbulence, we do not expect our conclusions on sensitivity to feedback to be altered.  
 

\section{Conclusions}\label{conclude}

We investigated the sensitivity of fifteen commonly applied turbulent statistics to the presence of stellar feedback. The goal of our analysis was to identify whether any of the statistics provide a robust indication of feedback: a smoking gun. Our parameter study was based on magneto-hydrodynamic simulations performed by OA15 with varying magnetic field strengths and degrees of feedback from stellar winds. We first post-processed the simulations with a radiative transfer code to produce synthetic $^{12}$CO(1-0) emission cubes. We then computed fifteen statistical metrics using the python package {\sc turbustat} (K16) and assessed the relative response of each statistic to changes in evolutionary time, magnetic field strength, and stellar mass-loss rate. We focused on those statistical formulations identified by K16 to respond to physical changes in parameters but were insensitive to noise fluctuations  and viewing angle: intensity PDF, skewness, kurtosis, power spectrum, PCA, SCF, bispectrum, VCA, VCS, $\Delta$-variance, wavelet transform, Genus, Cramer, number of dendrogram features and histogram of dendrogram feature intensities.  We illustrated each statistic via a comparison between a purely turbulent output and an output with identical turbulence but with embedded stellar sources launching winds (\S3). 

We then computed the distance metric, as defined for each statistic by K16, for each pair of outputs (\S4). This allowed us to both quantify changes and simplify the comparison by reducing each pair of data cubes to one characteristic number. We discovered that a variety of statistics exhibit sensitivity to feedback, and we present the following conclusions:
\begin{itemize}
\item The intensity PDF, skewness and kurtosis statistics are each sensitive to the degree of feedback, with strong wind models exhibiting very different distances than weak wind models. Skewness and kurtosis show sensitivity to evolutionary time to a lesser degree, while none are sensitive to magnetic field strength.
\item PCA shows strong sensitivity to wind strength and weak sensitivity to time evolution. In particular, the covariance matrix exhibits strong peaks at the characteristic wind shell expansion velocity ($v\sim 1-2\kms$), which we predict will be visible in observational data and could be a good diagnostic for wind-driven shell characterization.
\item  The SCF exhibits a strong response to feedback, which manifests as a statistically steeper SCF spectrum slope when wind feedback is present.
\item Both the Cramer statistic, which measures the spread of the variance, and bicoherence are strongly sensitive to feedback but mainly in a binary way. The Cramer distance is insensitive to the overall mass-loss rate and evolutionary time; however, of all the statistics it was the more purely sensitive to the magnetic field strength.
\item The SPS displays little sensitivity to feedback aside from an overall offset. No characteristic break appears to mark the energy injection scale. The SPS did show sensitivity to time evolution.
\item The VCA exhibits a weak response to both feedback and time evolution.
\item The VCS function shows a distinct signature of feedback. The transition between the density and velocity-dominated parts of the VCS curve occurred at higher velocities and larger scales in the case with winds. This suggests that the breakpoint may encapsulate information about the characteristic scale of feedback. The location of this point depends upon other cloud properties, such as optical depth and the velocity dispersion, however. Thus, VCS may be best applied to compare cloud sub-regions. The VCS also displays a weak response to magnetic field strength.
\item The bicoherence exhibits less correlation between scales in the case with feedback, which may be the result of the shells reducing magnetic wave propagation and scale-coupling. However, past work has demonstrated the bicoherence is also sensitive to local conditions, including the sonic and Alf\'venic Mach number, which may make absolute identification of feedback challenging. 
\item The wavelet transform and $\Delta$-variance display some response to the presence of feedback, although the 
degree of difference might not be remarkable in comparisons between observational datasets. The  wavelet transform also shows some sensitivity to time evolution and magnetic field strength.
\item The Genus statistic, which reflects the relative number of peaks and voids,  shows sensitivity to feedback at small scales: the number of voids declined when feedback was included. However, the effect was subtle and the color-plots comparing all pairs did not show strong trends.
\item Both dendrogram statistics show sensitivity to feedback. In the presence of feedback, the number of features follows a power-law for a much larger range of scales when feedback is present, rather than falling off steeply as in the purely turbulent case. Prior studies have found that power-law behavior does not occur for any cloud Mach number or magnetic field strength for purely driven turbulence. This suggests the number of features statistic may be a true scale-free metric, which could be used to identify and characterize feedback. The histogram of leaf intensities was broader in the case with feedback, which reflects the larger range of intensities associated with the increased temperatures and densities found in shells. 
\end{itemize}

In conclusion, our search for a smoking gun has yielding promising leads. Several statistics show clear features and variations associated with feedback that do not occur in purely turbulent simulations or in self-gravitating, turbulent simulations (as in K16) across a broad range of physical conditions. On the basis of these results, we recommend follow-up observational studies focusing on active star-forming regions utilizing the PDF, PCA, SCF, VCS and dendrograms. In particular, PCA is promising since it displays the characteristic velocity scale of the feedback.

Although these results provide motivation for optimism, we note several caveats.  We caution that many of the statistics have two or more distinct definitions in the literature. Our conclusions hold only for the definitions specified in K16; additional studies are needed to check alternative statistical conventions. Our simulations neglect gravity, which should be considered in future work. Finally, we note that many of the statistics are sensitive to the line opacity, 
and tracers with different optical depths and chemistry may yield different results. 

\acknowledgements We acknowledge helpful discussions with Blakesley Burkhart, Brandt Gaches, and Dinesh Kandel.  We thank Paul Clark for helpful comments, which improved the manuscript. RB received support from the Massachusetts Space Grant Consortium, EWK and EWR are supported by a Discovery Grant from the Natural Sciences and Engineering Research Council Canada (RGPIN-2012- 355247) and SSRO acknowledges support from NSF grant AST-1510021. Analysis was expedited by the helpful staff and facilities of the Massachusetts Green High Performance Computing Center in Holyoke, MA.

\bibliography{converted_to_latex.bib}

\appendix
\begin{figure}[h!]
\begin{center}
\includegraphics[width=0.7\columnwidth]{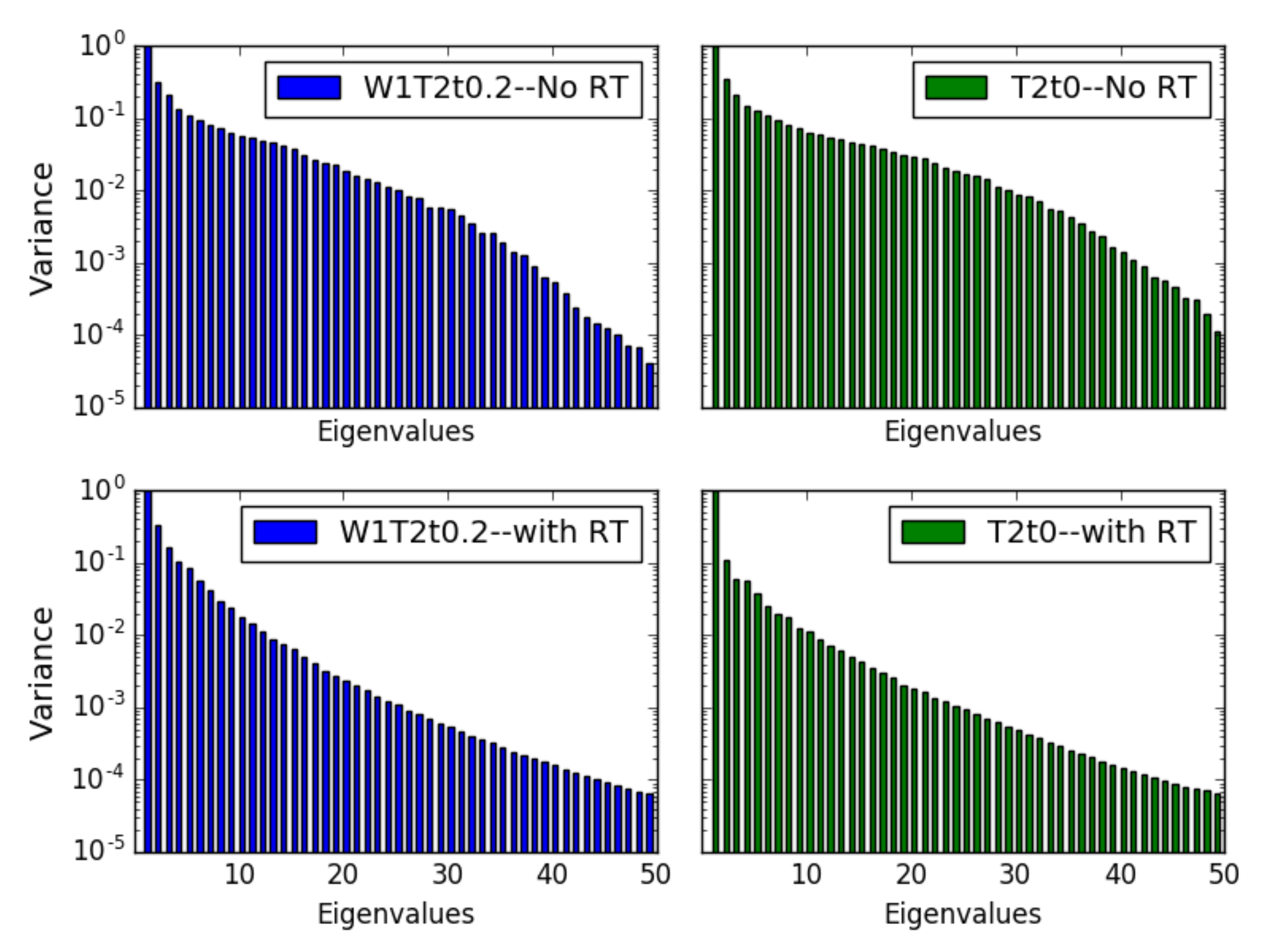}
\caption{\selectlanguage{english}The first 50 covariance matrix eigenvalues for runs W1T2t0.2 and T2t0 with and without the CO radiative transfer step. Output W1T2t0.2 is denoted by the color blue, while output T2t0 is denoted by the color green. The subplot labels indicate which plots include the radiative transfer step in the corresponding statistical measurement: ``No RT" shows the output for a position-position-velocity cube, and ``With RT" shows the output for a CO Spectral Cube. The normalized PCA distance metric yields  0.255 for our fiducial comparison with RT and 0.051 for that without RT.}
\label{eigvals_ppv_compare2}
\end{center}
\end{figure}

\begin{figure}[h!]
\begin{center}
\includegraphics[width=0.7\columnwidth]{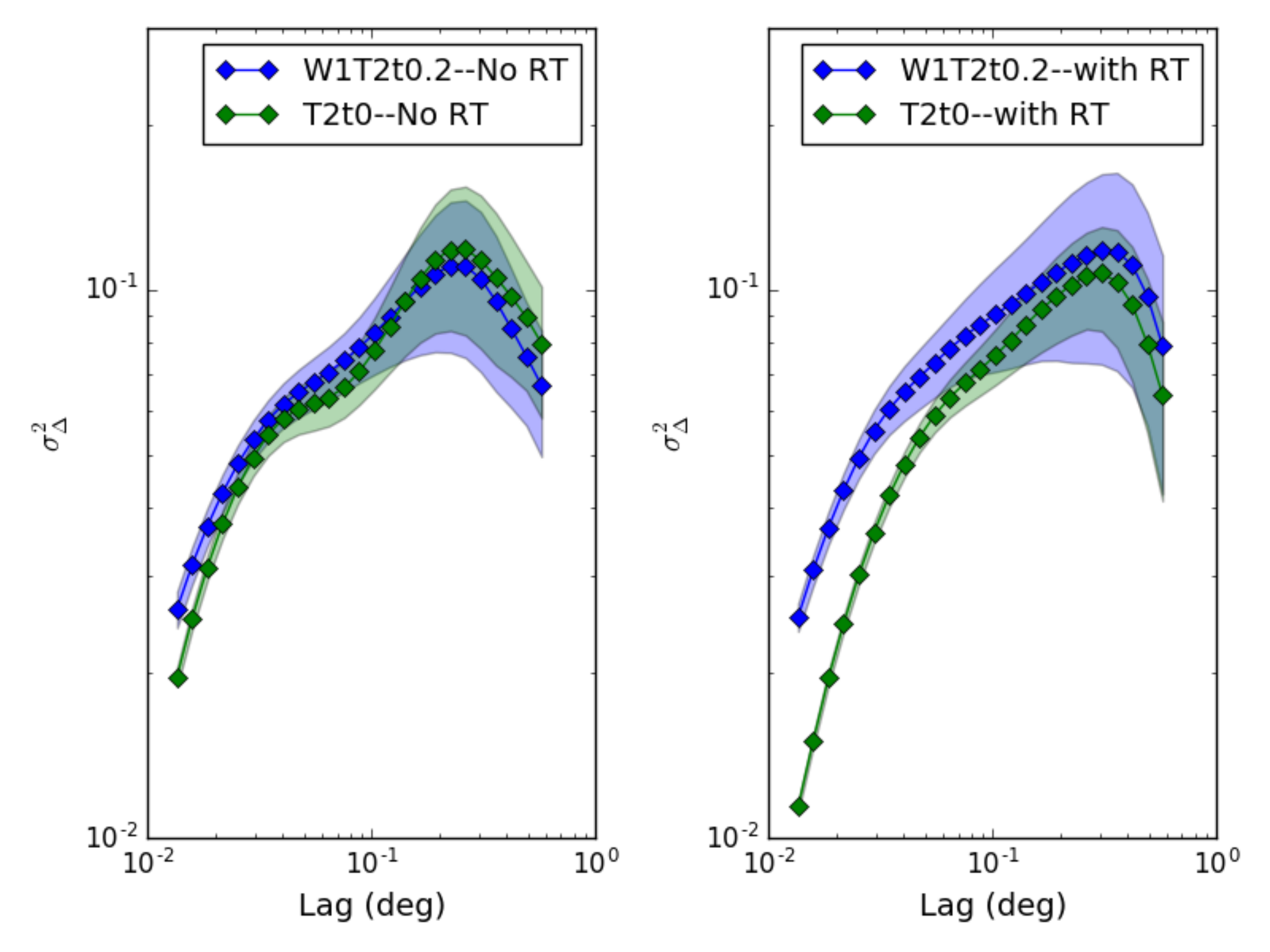}
\caption{\selectlanguage{english}The $\Delta$-variance spectra for outputs W1T2t0.2 (blue) and T2t0 (green) with (left) and without (right) the radiative transfer step. The $\Delta$-variance distance metric yields 0.79
for our fiducial comparison with RT and 0.30 for that without RT.}
\label{delvar_compare}
\end{center}
\end{figure}

In this appendix, we briefly assess the impact of the CO- radiative transfer (RT) step on our statistics. To do this, we compare the results from the CO spectral cubes to the results of raw position-position-velocity (PPV) cubes. The PPV cubes contain mass values instead of CO intensity values, and to analyze them, we perform the same statistical measurements that were done on the CO cubes in \S \ref{comparisons}. The specific methodologies for each statistic are also identical. 

We present the results for the PCA eigenvalues and the $\Delta$-variance in Figures \ref{eigvals_ppv_compare2} and \ref{delvar_compare}, respectively. 
Here, we find that RT provides an overall ``stretch," which actually highlights wind activity. Without RT, the statistical outputs for W1T2t0.2 and T2t0 appear nearly identical. 

The $\Delta$-variance statistic yields distinct outputs for W1T2t0.2 and T2t0 with RT, which are shown in Figure \ref{delvar_compare}. The distance metric yields 0.79 for the case with RT and 0.30 for the same two outputs without RT, less than half the RT value. Visually the shapes of the $\Delta$-variance for the non-RT outputs are also more similar. Comparing to the distances in Figure 17, we conclude feedback does not leave a significant figure in the raw PPV data.

RT also alters the shape of the PCA eigenvalue distribution, which easily distinguishes W1T2t0.2 from T2t0. Figure \ref{eigvals_ppv_compare2} shows the first 5 eigenvalues of W1T2t0.2 become noticeably greater than those of T2t0; without RT, they appear to be the same.
In terms of our distance metrics, the normalized PCA distance metric is 0.255 for the two outputs with RT and 0.051 for those without.  Because of the metric normalization, PCA shows that without RT the outputs are nearly identical to one another, while the RT outputs with and without feedback are distinct (see also the colorplot shown in the Figure 16 middle right panel). 

\begin{figure}[h!]
\begin{center}
\vspace{0.1in} 
\includegraphics[width=0.49\columnwidth]{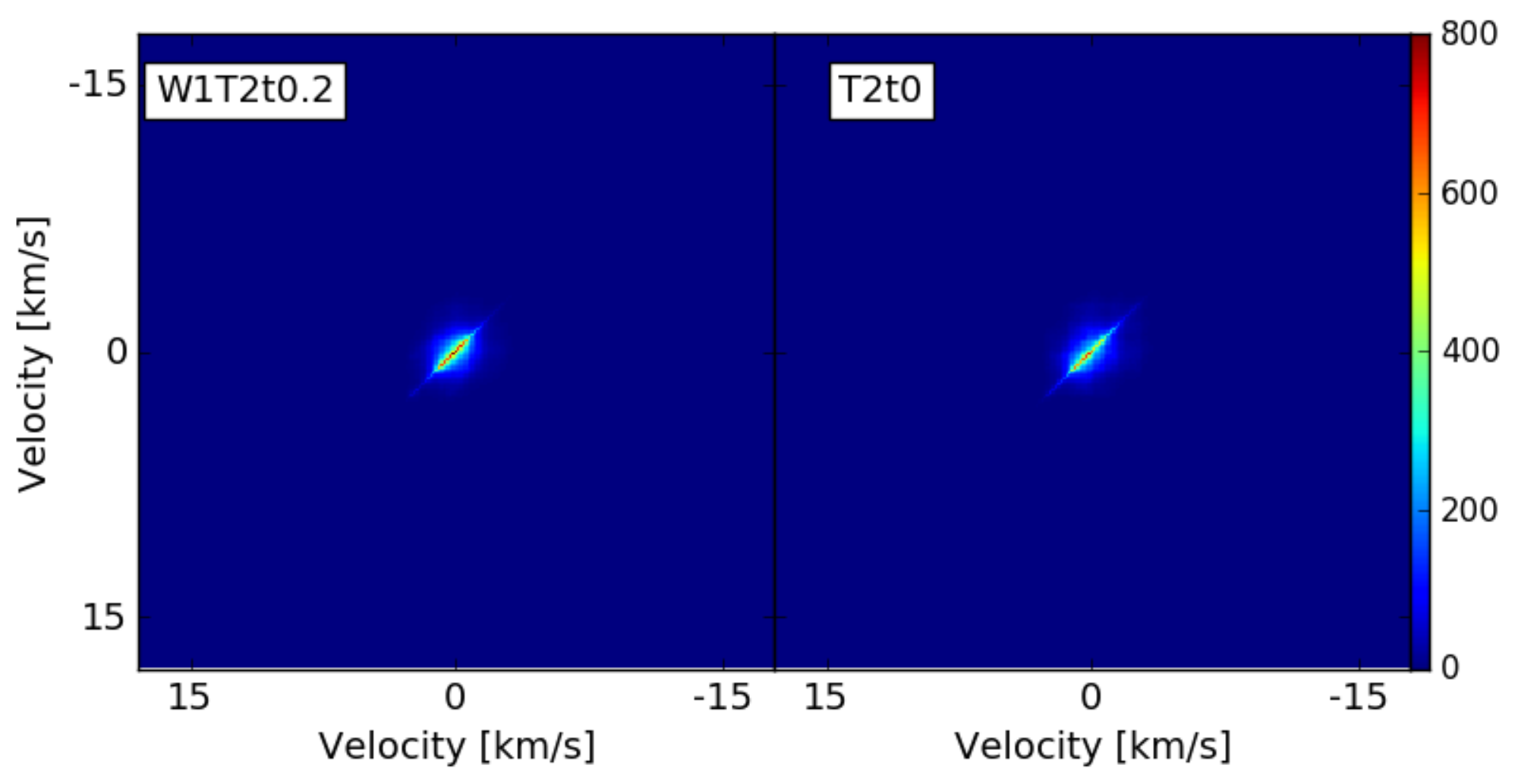} 
\vspace{0.1in} 
\includegraphics[width=0.49\columnwidth]{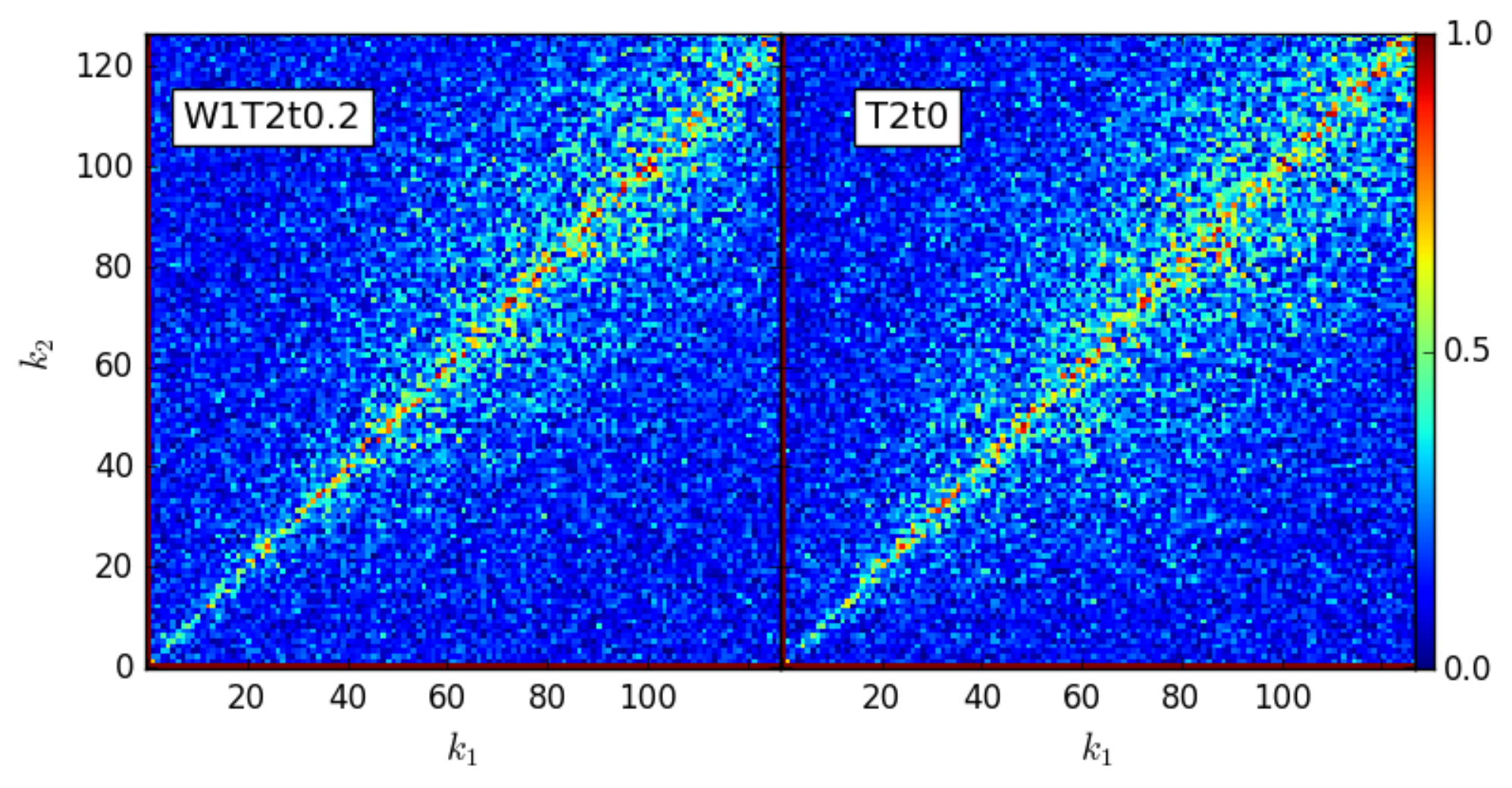}  
\caption{Left: The PCA covariance matrix for runs W1T2t0.2 (left) and T2t0 (right) without the CO radiative transfer step. The normalized PCA distance metric yields a distance of 0.051 for this pairing. Right: The bicoherence matrices for W1T2t0.2 (left) and T2t0 (right) without the radiative transfer step. The bicoherence distance metric produces a distance of 0.012 for this pairing.} \label{ppvplots1}
\end{center}
\end{figure}
\begin{figure}[h!]
\begin{center}
\vspace{0.1in} 
\vspace{0.1in} 
\includegraphics[width=0.50\columnwidth]{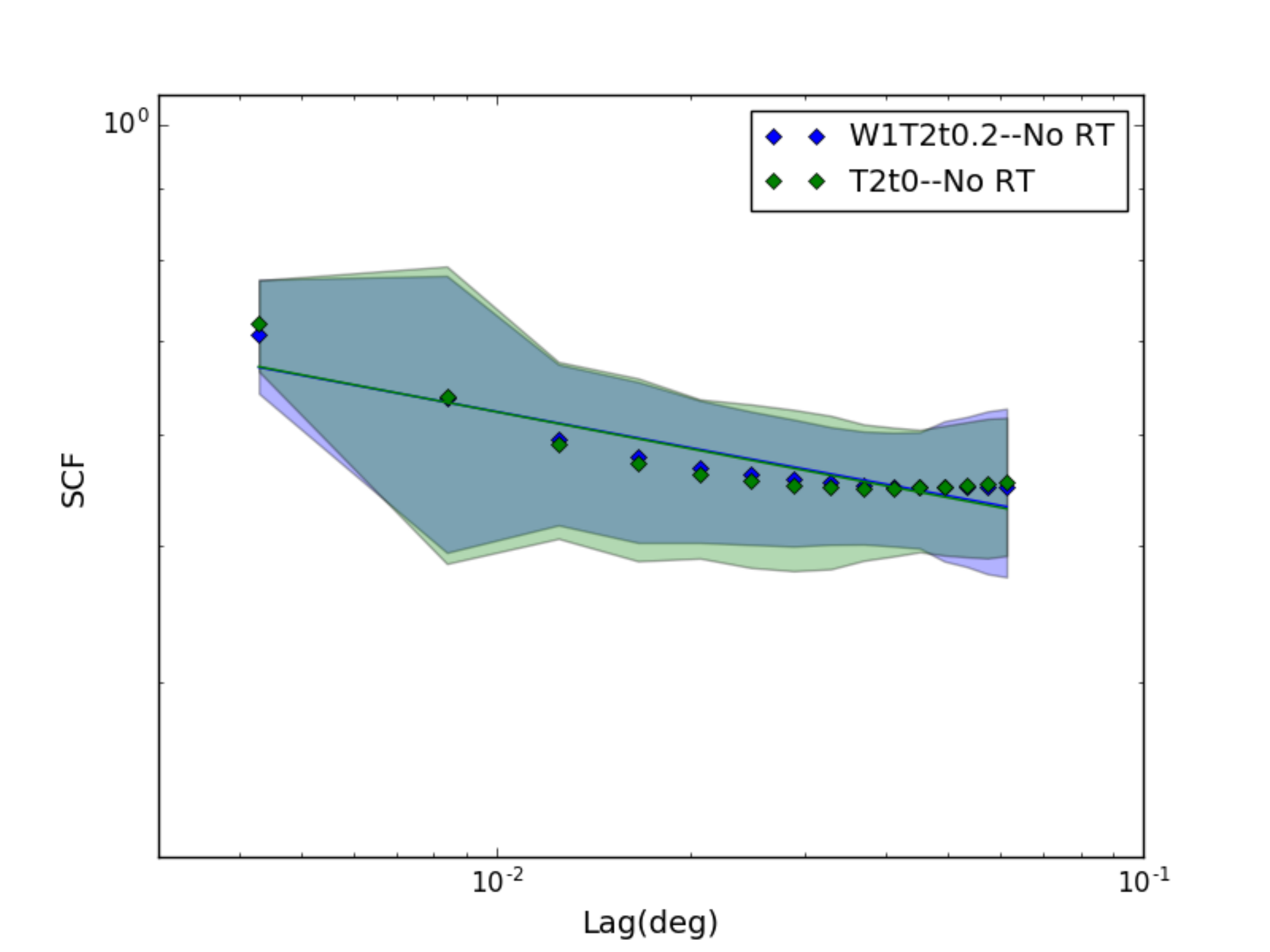} 
\vspace{0.1in} 
\includegraphics[width=0.47\columnwidth]{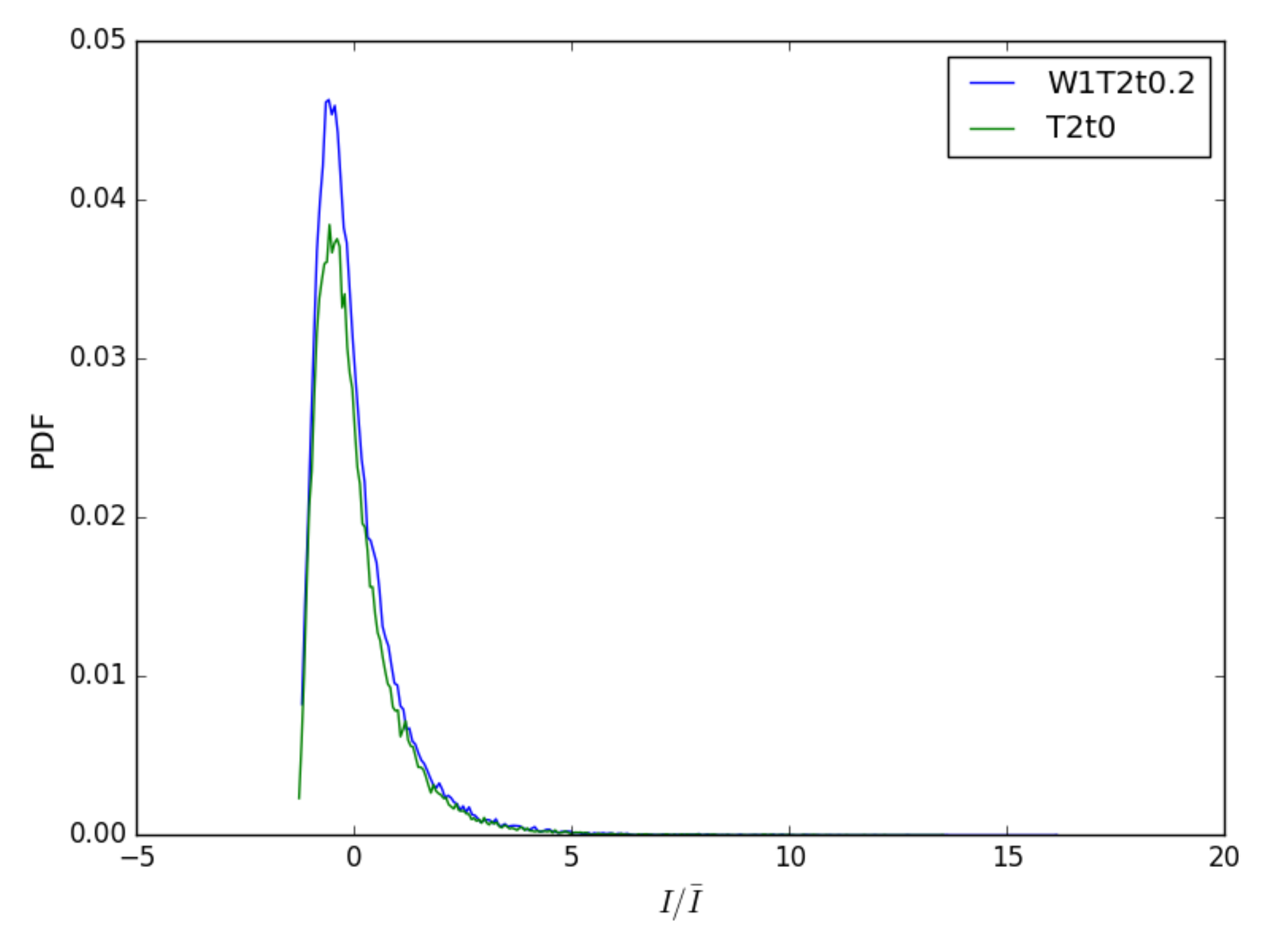}
\caption{ Left: The SCF spectrum for W2T2t0.2 (blue) and T2t0 (green) without radiative transfer. For this simulation pair, the SCF distance metric yields 0.005. Right: The PDFs for runs W2T2t0.2 (blue) and T2t0 (green) without radiative transfer. The PDF (Hellinger) distance yields 0.038.} \label{ppvplots2}
\end{center}
\end{figure}

The impact of RT is prominent in the most significant results in \S \ref{comparisons}. In \S\ref{comparisons} and \S4 we identified, quantified, and discussed the importance of the signatures that we identified from these statistics. In Figures \ref{ppvplots1},  \ref{ppvplots2}, and \ref{dendroppv} we show the PCA covariance matrices, bicoherence matrices, PDFs, SCF spectra, and dendrogram statatistics for the two outputs without RT. These are the statistics that previously exhibited large differences in their shapes and behavior in the presence of winds and appeared most promising as diagnostics. Similar to the comparisons is Figures \ref{eigvals_ppv_compare2} and \ref{delvar_compare}, the impact of winds disappears without radiative transfer.  

These combined results underscore that CO is an excellent observational tracer of stellar winds and outflows \citep[e.g.,][]{arce10,nakamura11}. Although stellar feedback is the key varying parameter between W1T2t0.2 and T2t0, without radiative transfer the statistics fail to respond to its influence. This is likely because CO emission saturates due to high optical-depth in high-density regions. The RT also boosts wind-related emission, which is warmer, and cuts out low-density gas, which contains little CO or is not strongly collisionally excited. Consequently, the dynamic range of the CO emission is small and favors emission arising from CO within shells. In contrast, without RT all the gas is weighted equally. Since the dynamic range of the gas density in the simulations spans 10 orders of magnitude, the same statistics without RT are biased towards high-density material, while, at the same time, the shell material is not enhanced by its temperature.



\begin{figure}[h!]
\begin{center}
\vspace{0.1in} 
\vspace{0.1in} 
\includegraphics[width=0.49\columnwidth]{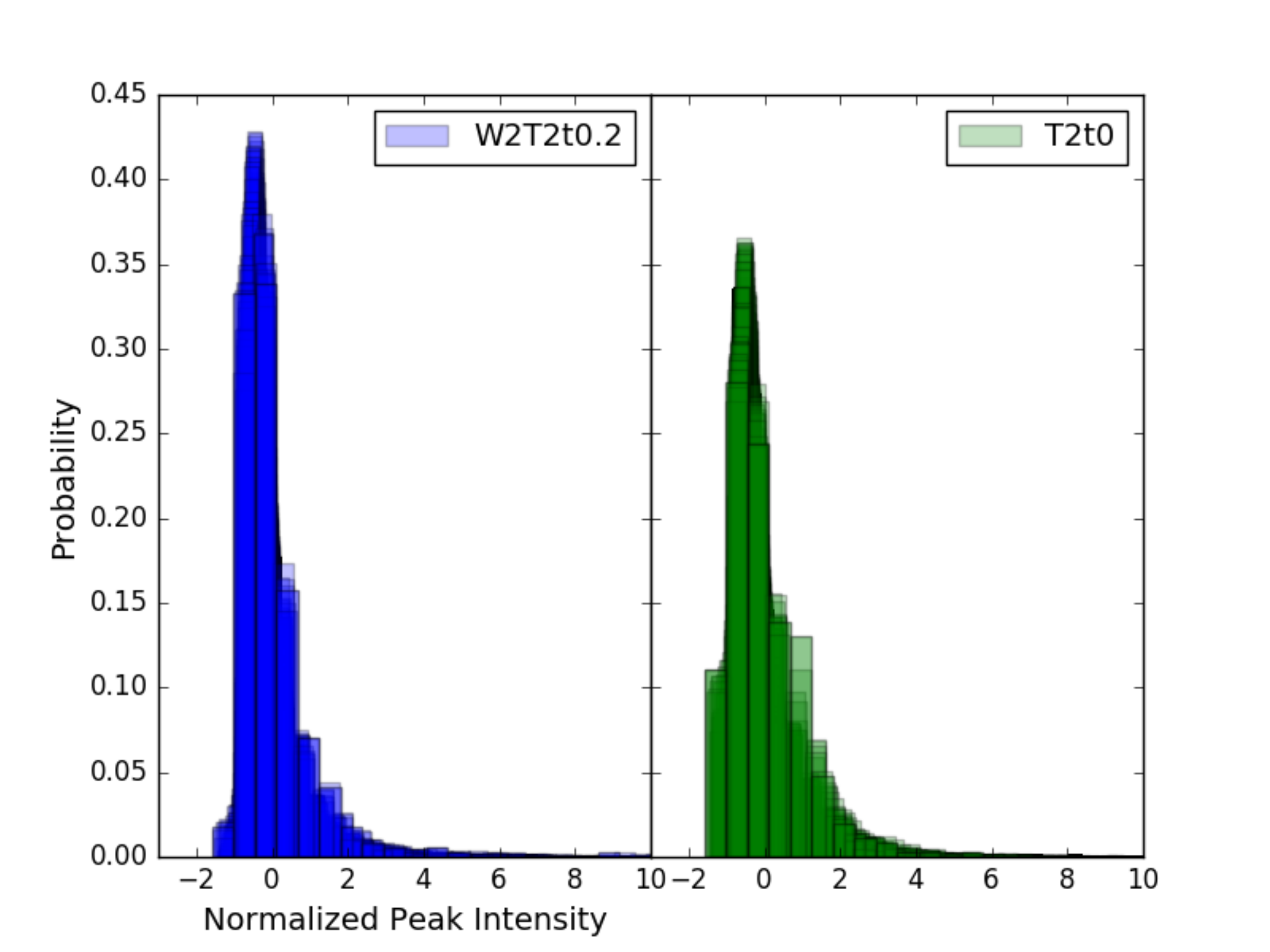} 
\vspace{0.1in} 
\includegraphics[width=0.48\columnwidth]{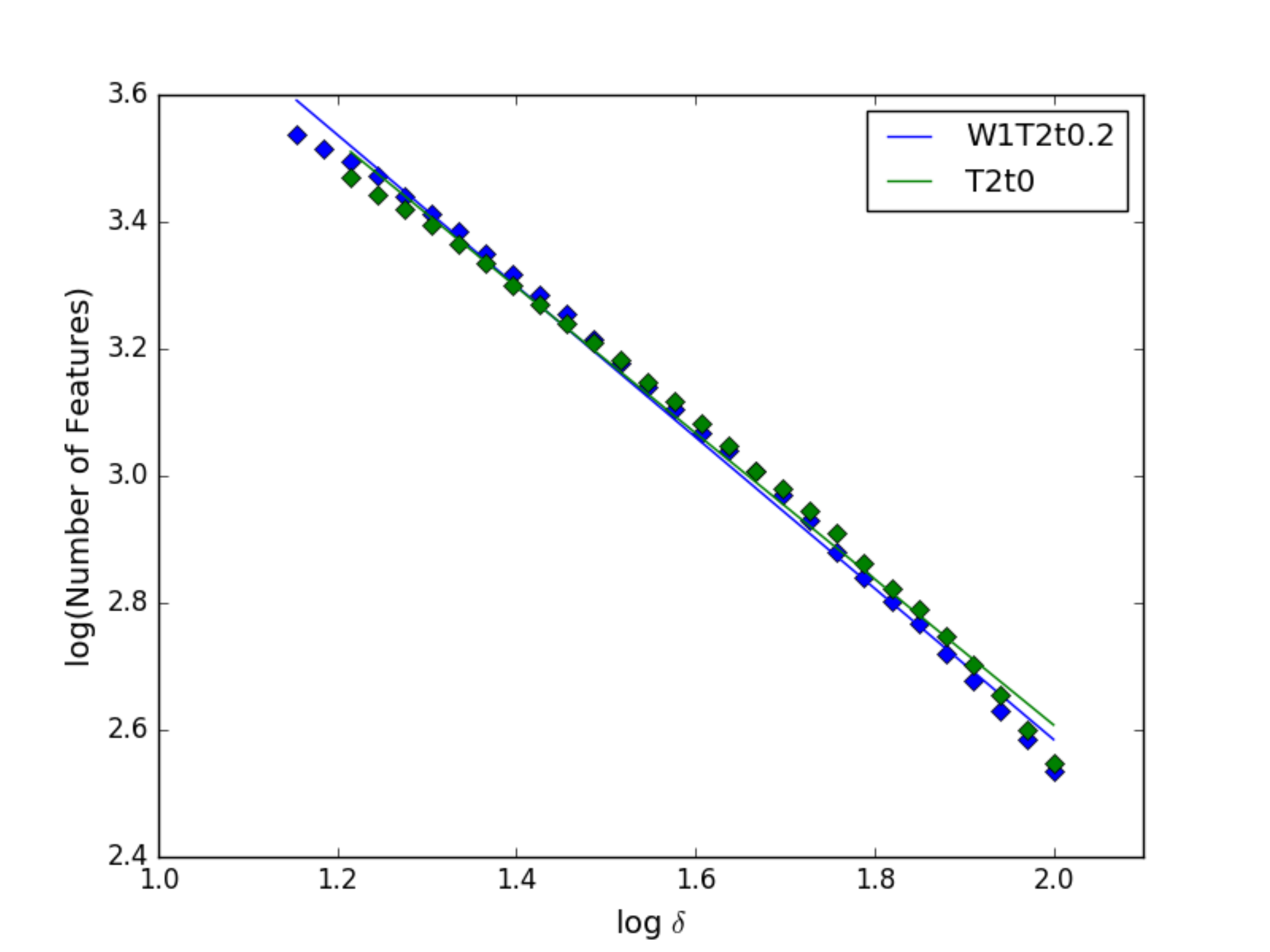}
\caption{Left: Histograms of the normalized dendrogram peak leaf intensities for the fiducial outputs without radiative transfer. The distance between the outputs is 0.103. Right: Number of dendrogram features statistic for two outputs without radiative transfer. Output W1T2t0.2 has a fitted slope of -1.19$\pm$0.02, while output T2t0 has a slope of -1.15$\pm$0.02. The distance metric yields 1.56.
 } \label{ppvdendro}
\end{center}
\end{figure}

\end{document}